\documentclass[sigconf]{acmart}

\usepackage[T1]{fontenc}
\usepackage[utf8]{inputenc}
\usepackage{lmodern} 
\usepackage{booktabs} 

\usepackage{pifont} 
\usepackage{graphicx}
\usepackage{enumitem}

\title{Sakshm AI: Advancing AI-Assisted Coding Education for Engineering Students in India Through Socratic Tutoring and Comprehensive Feedback}

\author{Raj Gupta }
\email{raj21410@iiitd.ac.in}

\affiliation{%
  \institution{IIIT Delhi }
  \city{New Delhi}
  \country{India}
}
\author{Harshita Goyal}
\email{p20240431@pilani.bits-pilani.ac.in}
\affiliation{%
  \institution{BITS Pilani}
  \city{Pilani}
  \country{India}
}

\author{Dhruv Kumar}
\email{dhruv.kumar@iiitd.ac.in}
\affiliation{%
  \institution{BITS Pilani and IIIT Delhi}
    \city{Pilani}
  \country{India}
}

\author{Apurv mehra }
\email{apurv@blendnet.ai}

\affiliation{%
  \institution{BlendNet AI }
  \city{Bengaluru}
  \country{India}
}

\author{Sanchit Sharma }
\email{sanchitsharma@blendnet.ai}

\affiliation{%
  \institution{ BlendNet AI  }
  \city{Pilani}
  \country{India}
}

\author{Kashish Mittal }
\email{kashish@blendnet.ai}

\affiliation{%
  \institution{BlendNet AI }
  \city{Bengaluru}
  \country{India}
}

\author{Jagat Sesh Challa}
\email{jagatsesh@pilani.bits-pilani.ac.in}

\affiliation{%
  \institution{BITS PILANI }
  \city{Pilani}
  \country{India}
}

\begin{document}

\begin{abstract}

The advent of Large Language Models (LLMs) is reshaping education, particularly in programming, by enhancing problem-solving, enabling personalized feedback, and supporting adaptive learning. Existing AI tools for programming education struggle with key challenges, including the lack of Socratic guidance, direct code generation, limited context retention, minimal adaptive feedback, and the need for prompt engineering.
To address these challenges, we introduce Sakshm AI, an intelligent tutoring system for learners across all education levels. It fosters Socratic learning through Disha, its inbuilt AI chatbot, which provides context-aware hints, structured feedback, and adaptive guidance while maintaining conversational memory and supporting language flexibility.
This study examines 1170 registered participants, analyzing platform logs, engagement trends, and problem-solving behavior to assess Sakshm AI’s impact. Additionally, a structured survey with 45 active users and 25 in-depth interviews was conducted, using thematic encoding to extract qualitative insights. Our findings reveal how AI-driven Socratic guidance influences problem-solving behaviors and engagement, offering key recommendations for optimizing AI-based coding platforms.
This research combines quantitative and qualitative insights to inform AI-assisted education, providing a framework for scalable, intelligent tutoring systems that improve learning outcomes. Furthermore, Sakshm AI represents a significant step toward Sustainable Development Goal 4: Quality Education, providing an accessible and structured learning tool for undergraduate students, even without expert guidance. This is one of the first large-scale studies examining AI-assisted programming education across multiple institutions and demographics.

\end{abstract}

\keywords{AI, Large Language Models, Sakshm AI, Education, E-learning}

\ccsdesc[300]{Human-centered computing~Empirical studies in HCI}

\maketitle

\section{Introduction}\label{sec:introduction}
The rapid advancements in AI technology herald a new era where artificial intelligence plays a transformative role in various aspects of daily life, including education \cite{denny2023computingeducationeragenerative}. Cutting-edge models such as OpenAI's GPT-3.5 \cite{openai2023} and GPT-4 \cite{openai2024}, Google’s Gemini \cite{google2023}, and Meta’s Llama-3 \cite{meta2023} possess the capability to generate human-like text, engage in critical and logical thinking, and facilitate a wide range of educational tasks \cite{farrokhnia2024swot}. These advancements have paved the way for LLMs to create resources such as programming exercises, code explanations, and model solutions, demonstrating their immense potential in supporting learning processes and academic endeavors.LLMs have the potential to transform education by enabling personalized learning experiences, fostering critical thinking, enhancing problem-solving skills, and supporting inclusive and adaptive learning strategies across diverse educational levels and disciplines \cite{kasneci2023chatgpt,pankiewicz2023large,guo,farrokhnia2024swot}.

LLMs are indispensable across multiple fields due to their ability to manage big data, solve complex algorithms, and streamline processing tasks. Their integration into education further enhances teaching and learning by simplifying complex concepts and offering adaptive strategies \cite{thacker2020age,yan2024practical,denny2023computingeducationeragenerative}.

Increased integration into academic settings has raised ethical concerns, including risks of bias, misuse, and challenges to employment and societal integrity \cite{farrokhnia2024swot,dai2024bias,jiao2024navigating,xu2023llms}. These dual facets—immense educational potential alongside pressing ethical challenges—are particularly significant in fields like computing, where LLMs' ability to autogenerate solutions to programming tasks disrupts traditional methods of assessment and learning \cite{denny2023computingeducationeragenerative,becker2023programming}. As the use of LLMs becomes more widespread, it is crucial to examine their impact not only on enhancing learning experiences but also on preserving the integrity of academic practices and addressing their broader societal implications \cite{rane2023chatgpt,yan2024practical,Kazemitabaar_2024}.

While many studies have explored critical thinking chatbots, Socratic questioning methods, and LLM-based educational tools \cite{chinedu,denny2023computingeducationeragenerative,kargupta2024instructassistllmbasedmultiturn}, challenges remain in integrating dynamic, context-aware guidance with scalable and pedagogically effective learning strategies. Existing literature often focuses on fixed, assignment-specific models or tools prioritizing efficiency over fostering deeper learning \cite{Zhao2020APIHelperHJ,WaldenACF} , leading to reliance on direct solutions rather than the development of conceptual understanding. While these approaches have proven effective in specific scenarios, challenges persist in fully integrating immediate AI assistance with cultivating independent problem-solving skills, particularly in fostering Socratic learning and critical thinking.  A significant limitation is the lack of adaptive, multi-turn AI guidance that retains context and evolves with the learner, facilitating iterative problem-solving, abstract reasoning, or independent learning. Additionally, LLM-powered educational assistants often require explicit prompt engineering, creating a cognitive burden that shifts focus away from problem-solving.

To address these challenges, we present Sakshm AI \cite{sakshm_website}, an innovative platform designed to transform the coding education landscape. At its core is the Disha chatbot, which leverages large language models (LLMs) and employs a Socratic questioning framework to guide learners through iterative problem-solving. Unlike traditional tools that offer direct solutions or require explicit prompting, Disha dynamically adapts to a learner’s context, maintaining a conversational flow that aligns with the student’s progress. The platform integrates a comprehensive question bank with over 450 data structures and algorithms (DSA) problems, categorized by difficulty and tagged with topics and company-specific relevance. This feature empowers learners to practice effectively by focusing on areas most aligned with their goals, such as preparing for interviews with companies like Amazon \cite{AMAZON}, Google \cite{Google}, and Flipkart\cite{Flipkart}.
Additionally, the platform’s integrated coding environment offers real-time test case validation, custom test cases, and detailed performance analysis, creating a holistic ecosystem for learning and evaluation. Table \ref{tab:comparison} highlights the differences between Sakshm AI and existing tools, showcasing its strengths in Socratic guidance, contextual hints, and adaptive learning. Unlike other platforms, the Sakshm AI platform seamlessly integrates AI assistance with adaptive learning strategies, fostering independent problem-solving and conceptual understanding. This comparison is further discussed in Section 6.3.

\begin{table*}[h!]
\centering
\renewcommand{\arraystretch}{1.5} 
\setlength{\tabcolsep}{10pt} 
\begin{tabular}{@{}lccccc@{}} 
\toprule
\textbf{Feature}                  & \textbf{ChatGPT} & \textbf{LeetCode} & \textbf{CodeAid} & \textbf{Codehelp} & \textbf{Sakshm AI} \\
\midrule
AI Assistant                 & Limited        & \ding{55}         & Limited        & Limited         & \ding{51}       \\
Socratic Guidance                 & \ding{55}        & \ding{55}         & Limited        & \ding{51}         & \ding{51}       \\
Contextual Hints                  & \ding{55}        & \ding{55}         & \ding{51}        & Limited          & \ding{51}       \\
Prompt Friendliness         & \ding{55}       & \ding{55}            & Limited        & \ding{55}         & \ding{51}       \\
Feedback on Code Quality          &  Limited        & Limited            & Limited        & Limited          & \ding{51}       \\
Company-Specific Questions        & \ding{55}       & Limited            & \ding{55}                 & \ding{51}       \\
Topic-Wise Practice               & \ding{51}        & \ding{51}         & \ding{55}         & \ding{55}         & \ding{51}       \\
Free Complexity Analysis          & Limited        & Limited            & \ding{55}        & \ding{55}         & \ding{51}       \\
Code Execution          & \ding{55}       & \ding{51}            & \ding{55}        & \ding{55}         & \ding{51}       \\
\bottomrule
\end{tabular}
\caption{Comparison of Features Across AI-Driven Educational Platforms}
\label{tab:comparison}
\end{table*}

A standout feature of Sakshm AI \cite{sakshm_website} is its performance feedback system, which evaluates submissions on multiple dimensions, including correctness, efficiency, and code quality. What sets Sakshm AI \cite{sakshm_website}apart is its ability to balance AI assistance with pedagogical rigor. The platform enforces Socratic guardrails through Disha, ensuring learners engage thoughtfully with problems rather than relying on shortcuts. Furthermore, the platform’s multilingual support, curated learning sheets, and personalized guidance make it an inclusive tool for diverse learners, from beginners to advanced coders.

The Sakshm AI platform has seen adoption across a diverse range of institutions and independent learners, with 1170 registered participants. We conducted a large-scale quantitative analysis of user interactions to systematically evaluate their impact, drawing insights from platform logs, engagement trends, problem-solving patterns, and chat interactions. Additionally, we employed a mixed-methods approach to assess the efficacy of Sakshm AI’s Socratic chatbot, Disha, integrating structured surveys and in-depth interviews. The quantitative component involved a structured survey administered to 45 active users who had completed at least five coding problems on the platform, enabling data collection on engagement trends, satisfaction levels, and user challenges. For the qualitative component, we conducted 25 in-depth, semi-structured interviews with diverse participants to gain nuanced insights into their experiences, problem-solving strategies, and interactions with the Disha chatbot. Thematic analysis of interview transcripts provided a rich contextual understanding, complementing the quantitative survey findings. This methodological framework ensures a holistic assessment of Sakshm AI's design, functionality, and potential to transform coding education through innovative pedagogical approaches.

 This study aims to address the following research questions (\textbf{RQs}):
\begin{enumerate}
    \item \textbf{RQ1} \label{RQ1}: How does the Socratic approach in Sakshm's AI chatbot Disha enhance critical thinking compared to direct solution platforms like ChatGPT?
    \item \textbf{RQ2} \label{RQ2}: What features and feedback mechanisms make Sakshm AI effective for structured learning compared to other coding platforms?
    \item \textbf{RQ3} \label{RQ3}: How can the Sakshm AI platform balance AI assistance and human tutoring to optimize student learning outcomes?

\end{enumerate}

By addressing the research questions outlined above, this paper comprehensively analyzes the design and impact of AI-driven educational platforms like Sakshm AI in programming education. We leverage user interactions, quantitative metrics, and qualitative feedback insights to identify key considerations for designing practical pedagogical tools. This study highlights critical trade-offs in features like Socratic guidance, structured learning, and AI-human collaboration, offering generalizable recommendations for the development of AI-powered coding assistants that foster critical thinking and practical skills in learners. Nevertheless, Sakshm AI represents a significant step towards achieving Sustainable Development Goal 4: Quality Education by providing an intelligent learning tool that supports students in developing problem-solving and coding skills, even without expert guidance. To the best of our knowledge, this is one of the first large-scale studies with 1,170 participants, covering a diverse demographic across multiple institutions.

\section{Related Works}\label{sec:related_works}
The emergence of Large Language Models (LLMs) has revolutionized the field of education, offering new opportunities to enhance learning through personalized and interactive AI-driven tools. \cite{alhafni2024llmseducationnovelperspectives,10.1145/3688094,denny2023computingeducationeragenerative}. These advancements have sparked significant research interest, particularly in their application to programming education, where they enable dynamic, adaptive, and Socratic learning experiences \cite{10.1145/3587102.3588815,Kazemitabaar_2024, denny2023computingeducationeragenerative}.  

\subsection{\textbf{Chatbots}}
Using large language models (LLMs) in education through interactive chatbots is transforming the ed-tech landscape, driven by the growing demand for educational tools that leverage natural language interfaces on computers and mobile devices \cite{chinedu}. Recent studies by Katzemitabaar et al.\cite{Kazemitabaar_2024} explore the development and deployment of CodeAid, an LLM-powered programming assistant designed to provide conceptual guidance without revealing code solutions.  Similarly, Knill et al. \cite{knill} highlight that chatbot interactions enable educators to monitor the types of questions students ask, identify challenging areas in a subject, and assess student learning capabilities. Furthermore, As highlighted by Birillo et al. \cite{birillo}, large language models (LLMs) can effectively guide students by generating next-step hints through a chain-of-thought approach. Integrated into the JetBrains Academy plugin, their system helps students by breaking tasks into subgoals, generating code, and offering clear textual explanations within the learning environment. Chinedu et al. \cite{chinedu} developed Python-Bot the SnatchBot platform, which assists novice programmers in understanding Python’s syntax and semantics, effectively supporting their program comprehension. Zhao et al. \cite{Zhao2020APIHelperHJ} created APIHelper, which helps junior Android developers learn API usage by analyzing real-time calls. Likewise, Walden et al. \cite{WaldenACF} introduced a chatbot to teach secure programming in PHP. \emph{While CodeAid and other tools like Python-Bot, APIHelper, and secure programming chatbots provide fixed, assignment-specific support for particular languages and functionalities, Sakshm AI distinguishes itself by supporting multiple programming languages, enabling open-ended and context-aware queries, maintaining conversational context, offering detailed feedback on code quality and optimization, integrating Socratic questioning to promote critical thinking, and delivering scalable, personalized guidance tailored to individual student needs.}

Recent works emphasize the growing importance of learning to craft effective prompts for LLMs, a skill critical for generating accurate and context-aware outputs. Studies like Denny et al. \cite{denny2023computingeducationeragenerative} introduce innovative approaches, such as 'Prompt Problems,' to teach this skill effectively. Babe et al. \cite{babe2023studentevalbenchmarkstudentwrittenprompts} show that Code LLMs perform inconsistently with novice-written prompts, highlighting significant variability in prompting techniques and the importance of teaching effective, prompt formulation. Goparthi \cite{goparthi} mentions that prompt engineering enables users to craft precise inputs for LLMs, unlocking the full potential of generative AI by overcoming challenges. \emph{In contrast, Sakshm AI stands out by removing the necessity of learning intricate prompting techniques, a barrier often faced with traditional LLMs. Unlike platforms that require users to refine and adjust prompts for accurate outputs, Sakshm AI’s design ensures that students can focus entirely on learning and problem-solving rather than struggling with input formulation. Remembering the flow of the discussion offers a personalized and efficient learning experience that bridges the gap between AI assistance and human-like adaptability.}

\subsection{\textbf{Socratic Learning}}
Socratic questioning, named after the ancient Greek philosopher Socrates \cite{socrates_wikipedia}, is a methodical \cite{methodical_wiktionary} approach to asking questions that helps explore intricate concepts, uncover hidden assumptions, and distinguish between what is known and what is uncertain \cite{paul2019thinker}.  Socratic questioning encourages deeper analysis, enhances understanding, and promotes thoughtful inquiry and diverse perspectives \cite{zare2015use,favero2024enhancing,ho2023thinking}. Chatbots can facilitate the Socratic method by leading users through a series of thought-provoking, open-ended questions that stimulate critical thinking \cite{favero2024enhancing}.
Kargupta \cite{kargupta2024instructassistllmbasedmultiturn} shows that TreeInstruct leverages Socratic questioning to guide students in debugging code, dynamically adapting to their knowledge through a state-space planning algorithm. It outperforms baselines in handling single-bug and multi-bug scenarios, demonstrating its effectiveness as an instructor. Hossami et al. \cite{hossami} introduce a manually curated dataset of multi-turn Socratic advice designed to help novice programmers debug solutions to simple computational problems. \emph{In contrast, Sakshm AI's Disha chatbot excels in maintaining a structured Socratic approach that prioritizes logical reasoning and independent learning, distinguishing it from more algorithmically driven models like TreeInstruct \cite{kargupta2024instructassistllmbasedmultiturn}. Unlike Hossami et al.’s \cite{hossami} curated dataset approach, which may require extensive manual effort, Disha's predefined Socratic guardrails offer a scalable solution for fostering critical thinking across diverse user groups. Furthermore, Sakshm AI integrates contextual understanding, enabling it to provide guided feedback tailored to the user’s ongoing progress, which complements its role as an effective educational assistant beyond debugging scenarios.}

\subsection{\textbf{LLMS in Computer Education }}
LLMs have significantly impacted computer education and programming by generating and explaining code to support learning [1]. Their advanced ability to mimic human-like content has also improved tasks such as reading, writing, and coding across different educational levels \cite{prather2023transformed,denny2023computingeducationeragenerative}. Leading organizations such as OpenAI with its suite of models including ChatGPT \cite{openai2023}, Codex  \cite{openai_codex}, and GPT-x \cite{openai_gptx}, Amazon with CodeWhisperer \cite{aws_codewhisperer}, etc are actively advancing these capabilities. Moreover, innovations like GitHub Copilot \cite{github_copilot} and coding chatbots such as OpenAI Codex \cite{openai_codex}, and DeepMind AlphaCode \cite{deepmind_alphacode} illustrate the expanding role of LLMs in transforming the programming landscape \cite{greeff2023exploring}. Liffiton et al. \cite{liffiton2023codehelp} introduced CodeHelp, an LLM-powered tool that provides on-demand programming assistance without revealing direct solutions, thereby reducing student over-reliance. As shown in the paper by Ma et al. \cite{ma2024integrating}, RAGMan, an LLM-powered tutoring system, supports course-specific AI tutors using Retrieval Augmented Generation (RAG) to provide accurate, aligned responses, fostering a judgment-free learning environment that 78\% of students found beneficial to their learning. Leinonen et al. \cite{leinonen2023comparing} show that while student- and LLM-generated code explanations are similar in length, LLM explanations are seen as more accurate and easier to understand. They suggest that LLM-generated explanations can serve as helpful examples for students, especially in early learning stages. Denny et al. \cite{denny2023computingeducationeragenerative} highlight the challenges and opportunities that code generation models present for computing educators, particularly in introductory programming classrooms and their potential impact on pedagogical practices. Kasneci \cite{kasneci2023chatgpt} shows that large language models can enhance education but require critical thinking, fact-checking, and awareness of biases for responsible use. \emph{ In contrast, Sakshm AI integrates LLM capabilities with a Socratic-guided approach, ensuring students engage critically with problems rather than relying solely on direct solutions. Unlike traditional LLM-powered tools, which may prioritize speed and convenience, Sakshm AI focuses on contextual feedback and structured learning, promoting deeper understanding and independent problem-solving. By balancing AI assistance with guided inquiry, Sakshm AI mitigates over-reliance risks and fosters a more active and thoughtful learning experience.}

\section{Sakshm AI System Design}\label{sec:design}

\subsection{Application Architecture}
The Sakshm AI platform is designed as a scalable and modular system that integrates various AI-driven components to enhance coding education. The architecture follows a microservices-based approach \cite{bakshi2017microservices,pereira2021security,bogner2017automatically}, ensuring flexibility and efficient handling of user interactions. The core system comprises the following components:

\begin{itemize}
    \item \textbf{Frontend:} A web-based user interface built with JavaScript frameworks to provide an interactive learning experience.
    \item \textbf{Backend:} A robust backend developed using Python and Flask, responsible for managing user requests, processing AI responses, and interfacing with the database.
    \item \textbf{Database:} A relational database storing user attempts, feedback, chat interactions, and evaluation results.
    \item \textbf{AI Models:} An ensemble of LLMs and other AI models is used for code evaluation, hint generation, and personalized feedback.
    \item \textbf{Orchestration Layer:} A middleware component that coordinates the interaction between different subsystems, ensuring efficient task management and response generation.
\end{itemize}

\subsection{LLM-Based Analysis Approach}
Sakshm AI employs Large Language Models (LLMs) to analyze and evaluate student code submissions. The LLM-based analysis involves:

\begin{itemize}
    \item \textbf{Code Understanding:} The LLM interprets student-submitted code to determine correctness, efficiency, and code quality.
    \item \textbf{Hint Generation:} Instead of directly providing answers, the model employs a Socratic approach, guiding students toward the correct solution through contextual hints.
    \item \textbf{Feedback Mechanism:} The AI generates structured feedback based on pre-defined rubrics and evaluates submitted code against pre-defined test cases, ensuring a comprehensive assessment that aligns with human grading practices.
    \item \textbf{Adaptive Learning:} The model adapts to user interactions, improving its recommendations over time based on past attempts and performance trends.
\end{itemize}

\subsection{Analysis Using Other AI Models}
Apart from LLMs, Sakshm AI integrates additional AI models to enhance its evaluation capabilities:

\begin{itemize}
    \item \textbf{Code Similarity Detection:} Leveraging transformer-based models to compare student code against reference solutions while detecting plagiarism and inefficient implementations.
    \item \textbf{Static Analysis Tools:} AI-driven static code analysis tools to identify syntax errors, logic flaws, and code smells.
    \item \textbf{Automated Test Execution:} A dynamic execution environment that runs test cases on submitted code to validate correctness and performance.
    \item \textbf{Sentiment Analysis:} AI-powered models assess student responses and engagement levels to optimize feedback delivery.
\end{itemize}

\subsection{Key Features of Sakshm AI Platform}

\subsubsection{\textbf{Question Bank}}  
\begin{figure*}
    \centering
    \includegraphics[width=\linewidth]{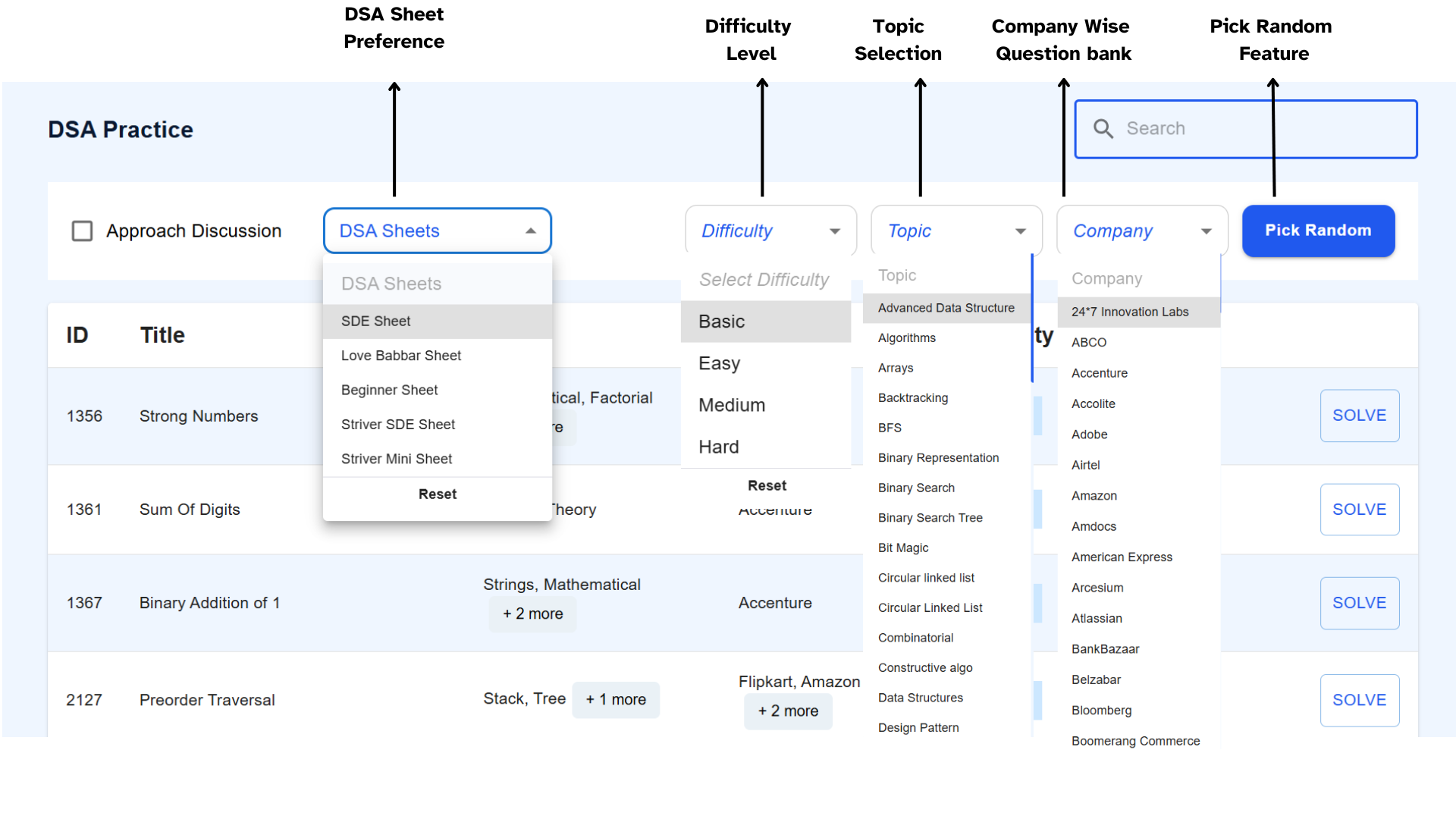}
    \caption{DSA Practice Features}
    \label{fig:dsa sheet}
\end{figure*}

\hfill\\
The system provides a comprehensive question bank comprising over 450 data structures and algorithms (DSA) problems. These questions are categorized by difficulty (basic, easy, medium, hard) and tagged with prerequisite topics and companies known to ask these questions. Students can filter the question set by difficulty, topic, or company, enabling them to target their practice effectively, as illustrated in Figure \ref{fig:dsa sheet}

\subsubsection{\textbf{Coding Interface with Integrated AI Assistance}}  \hfill\\

\begin{figure*}
    \centering
    \includegraphics[width=\linewidth]{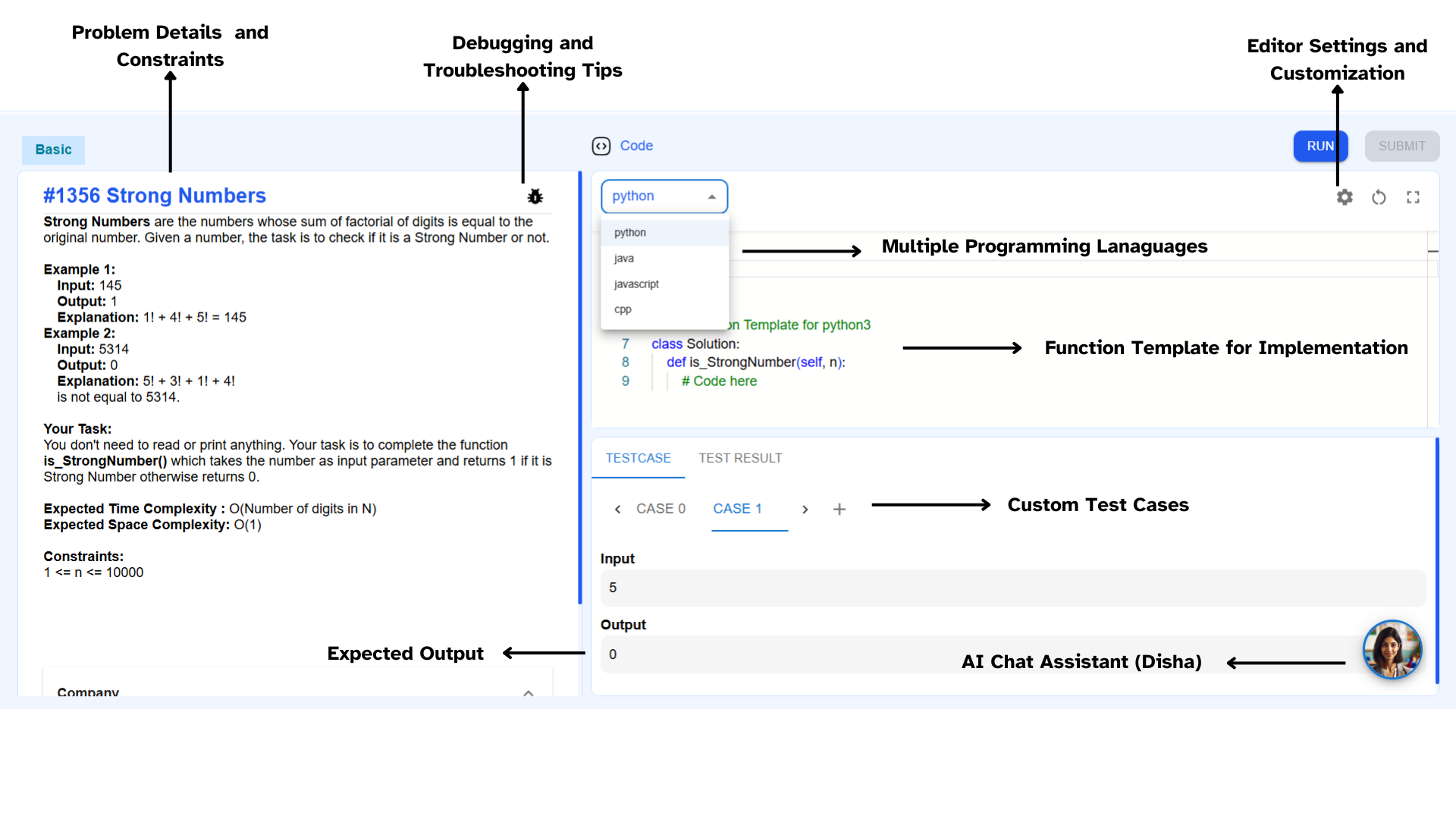}
    \caption{Integrated Coding Interface with AI Assistance}
    \label{fig:coding}
\end{figure*}

As shown in figure \ref{fig:coding}, the coding interface provides an integrated development environment (IDE) with features such as curated problem sheets, a customizable code editor, and test case validation to enhance the learning experience.

\paragraph{Question Selection Panel} 
When a student selects a question, they are directed to a dedicated coding workspace that includes:
\begin{itemize}
    \item \textbf{Display of the Chosen Question:} Shows constraints, expected input-output formats, and associated company tags.
    \item \textbf{Curated Sheets:} Includes sheets like a 38-question beginner’s sheet for progressive skill-building.
    \item \textbf{Global Search and Random Question Option:} Helps users discover problems efficiently.
\end{itemize}

\paragraph{Code Editor and Testing Environment} 
The platform offers a standard integrated development environment (IDE) supporting Python, Java, C++, and JavaScript. Features include:
\begin{itemize}
    \item \textbf{Customizable Code Editor:} Allows font, theme, and layout personalization.
    \item \textbf{Pre-defined Input Functions:} Helps students focus on implementing core logic.
    \item \textbf{Test Case Validation:} Displays passing cases in green and failing ones in red, with a summary of test performance.
    \item \textbf{Custom Test Cases:} Enables students to create and add their test cases.
\end{itemize}

\subsubsection{\textbf{Integrated AI Chat Assistant (Disha)}}  \hfill\\

\begin{figure*}
    \centering
    \includegraphics[width=\linewidth]{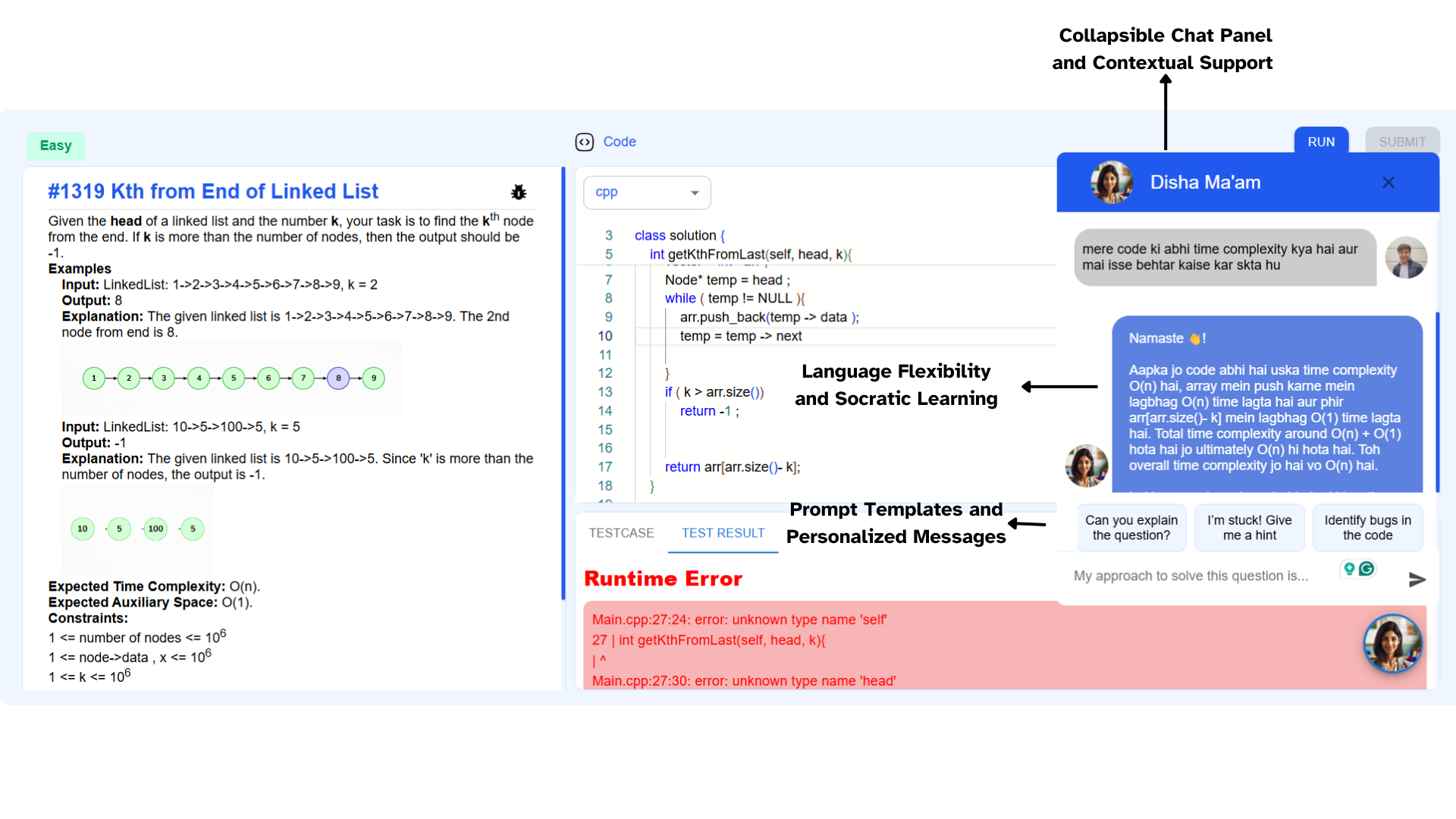}
    \caption{Disha AI Assistant Feature}
    \label{fig: disha}
\end{figure*}

As shown in figure \ref{fig: disha}, Disha, the AI assistant, enhances the coding experience by providing real-time support, analyzing user code, and offering contextual guidance. It encourages critical thinking, assists with debugging, and supports multiple languages. The following are the key features of Disha:

\begin{itemize}
    \item \textbf{Contextual Support : } 
A collapsible chat panel provides direct access to Disha, the AI assistant, which greets the user upon entering the coding environment. It draws on the user’s current code, the problem statement, and test case results to offer contextual guidance.
    \item \textbf{Socratic Method : } 
Disha employs a Socratic questioning style, encouraging critical thinking by prompting students to reason through challenges. It offers hints and strategies without directly revealing solutions.
    \item \textbf{Prompt Templates and Personalized Messages : } 
The assistant includes pre-set prompts such as “Can you explain the question?”, “I’m stuck, give me a hint,” and “Identify bugs in the code.” If user data is available, greetings and follow-ups are personalized for better engagement.

    \item \textbf{Language Flexibility and Snippet Support : } 
Disha supports multiple languages and scripts, assisting users in their native language. It provides concise code snippets for syntax-related queries to overcome specific coding hurdles.
\end{itemize}

\subsubsection{\textbf{Automated Performance Reports}} 
\begin{figure*}
    \centering
    \includegraphics[width=\linewidth]{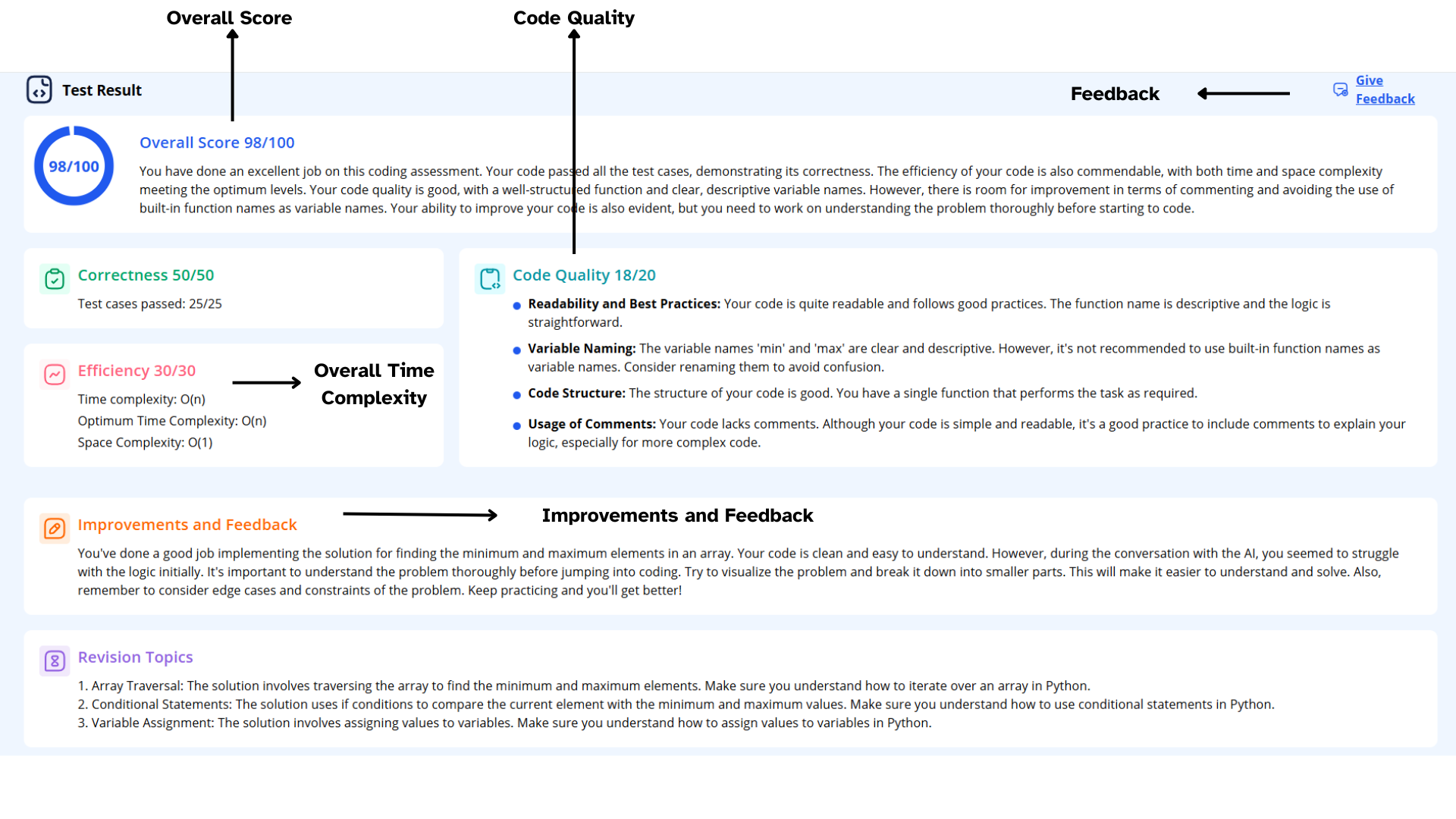}
    \caption{Performance Reports}
    \label{fig: feedback}
\end{figure*}

\hfill\\
As shown in figure \ref{fig: feedback}, Upon submission of a final solution, the system generates a detailed performance report, including:
\begin{itemize}
    \item \textbf{Correctness (50 points):} Based on the number of test cases passed.
    \item \textbf{Efficiency (30 points):} Time and space complexity evaluated by AI-driven analysis.
    \item \textbf{Code Quality (20 points):} Judged on readability, variable naming, structure, and commenting.
\end{itemize}

The total score out of 100 is accompanied by an AI-generated summary highlighting strengths, areas for improvement, and recommended topics for further study. This feedback loop ensures students receive actionable insights for skill development.

\subsubsection{\textbf{Leaderboard and Performance Insights}} 
\begin{figure*}
    \centering
    \includegraphics[width=\linewidth]{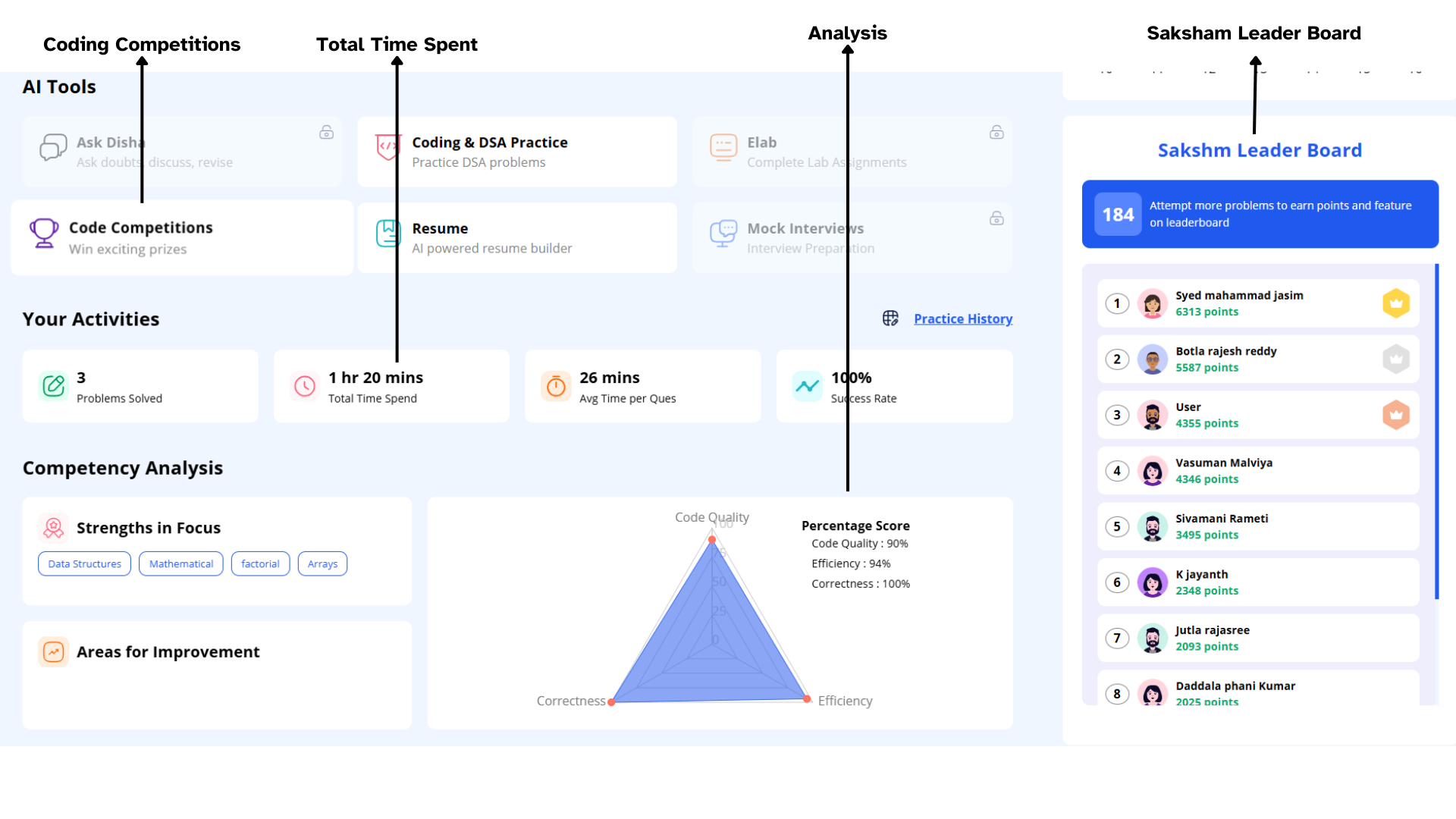}
    \caption{Leaderboard and Performance Insights}
    \label{fig: LeaderBoard}
\end{figure*}

\hfill\\
As shown in figure \ref{fig: LeaderBoard}, the platform provides in-depth performance tracking through features like coding competitions, detailed analysis, and the figure AI leaderboard, which ranks users based on total scores, efficiency, and problem-solving streaks. It also tracks the total time spent on coding, average time per question, and problem-specific insights, helping users optimize their approach. Additionally, AI-generated progress reports highlight strengths, areas for improvement, and recommended topics for further study, ensuring a data-driven and goal-oriented learning experience.

\subsection{Prompt Design for Effective Interaction}
Prompt engineering is critical in ensuring that LLMs provide meaningful and contextually appropriate responses. The design considerations include:

\begin{itemize}
    \item \textbf{Contextual Awareness:} Prompts are designed to retain session context, allowing AI to understand previous student interactions.
    \item \textbf{Step-by-Step Guidance:} AI responses follow a scaffolded learning approach, breaking down problem-solving into digestible steps.
    \item \textbf{Error Identification and Correction:} Tailored prompts help students debug their code by identifying potential errors and suggesting corrections.
    \item \textbf{Adaptive Prompting:} Based on the complexity of the student’s submission, the system dynamically adjusts the level of detail in hints and explanations.
\end{itemize}

\subsection{Workflow Description}
The operational workflow of the figure AI platform consists of the following steps:

\begin{enumerate}
    \item \textbf{User Submission:} Students submit their code solutions through the web interface.
    \item \textbf{Initial Analysis:} The system performs static analysis to detect syntax errors and structural issues.
    \item \textbf{AI Evaluation:} The LLM and other AI models assess the code for correctness, efficiency, and adherence to best practices.
    \item \textbf{Hint and Feedback Generation:} The AI generates contextual hints and detailed feedback, guiding students towards improvements.
    \item \textbf{User Interaction:} Students engage with the AI assistant, refining their code based on suggestions and explanations.
    \item \textbf{Final Assessment:} Upon successful revision, the system provides a final evaluation score and learning insights.
    \item \textbf{Progress Tracking:} The dashboard updates with new performance metrics, helping students monitor their progress over time.
\end{enumerate}

This structured workflow ensures a seamless learning experience, combining AI-powered evaluation with interactive tutoring to foster deeper understanding and coding proficiency.

\begin{figure*}[h]
    \centering
    \includegraphics[width=\linewidth]{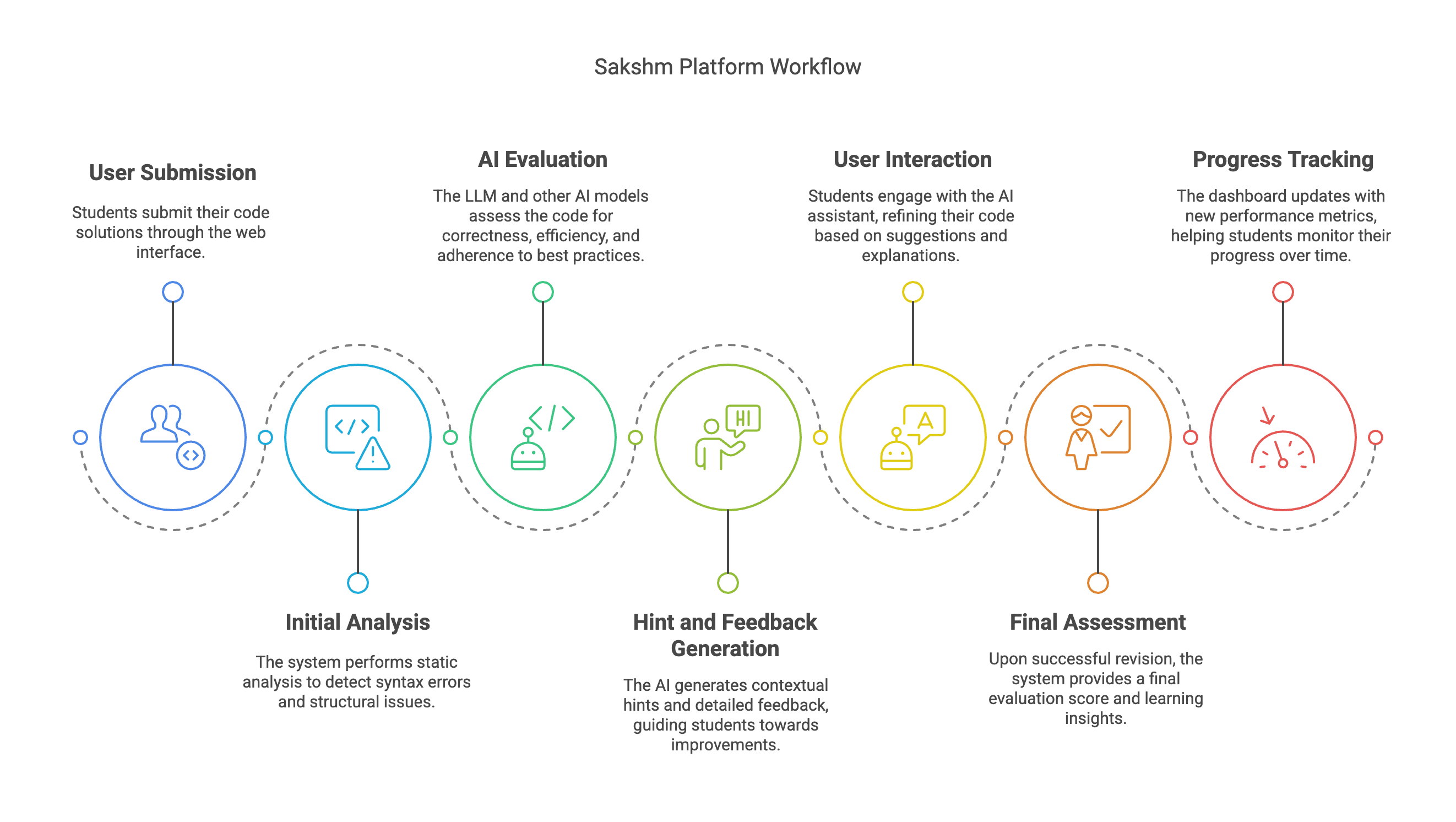}
    \caption{Leaderboard Snapshot }
    \label{fig:LeaderBoard}
\end{figure*}

\section{Methodology}\label{sec:methodology}
\subsection{Data Collection}

We employed a mixed-methods approach to investigate the usage and effectiveness of the Sakshm AI platform in supporting users' coding journeys. Data collection involved three primary techniques: user logs \cite{jansen2006search, bilodeau2010data}, surveys \cite{de2012choosing, couper2005technology}, and personalized interviews \cite{alshenqeeti2014interviewing, heath2018s}.

The Sakshm AI platform has 3,951 registered participants, as shown in the institute-wise distribution (see Table \ref{table:institute_breakup}). Among these, 1,170 participants have solved at least one coding problem on the platform. We conducted a large-scale quantitative analysis of 1,170 user interactions to systematically evaluate its impact, drawing insights from platform logs, engagement trends, problem-solving patterns, and chat interactions. For the survey component, we targeted a subset of users and surveyed 45 participants (see Table \ref{table:institute_breakup_survey}) who had attempted a minimum of five coding problems. A random sampling method was employed to select the survey respondents.

To ensure accessibility and maintain respondent anonymity, the survey was conducted via Google Forms, which did not require participants to provide personal information, such as names or phone numbers. The study aimed to gather insights into user experiences, problem-solving approaches, and interactions with the platform, including their engagement with the Disha AI chatbot. 

\begin{table}[t]
    \centering
    \small  
    \caption{\textbf{Registered Users Institute-wise Breakup}}
    \label{table:institute_breakup}
    \begin{tabular}{@{}lc@{}}
        \toprule
        \textbf{Institute} & \textbf{Number of Users} \\
        \midrule
        IIITD & 199 \\
        Ganpat University & 446 \\
        None & 2869 \\
        UPES & 92 \\
        SVCET & 205 \\
        BITS & 140 \\
        \bottomrule
    \end{tabular}
\end{table}

\begin{table}[t]
    \centering
    \small
    \caption{\textbf{Survey Responses Institute-wise Break up}}
    \label{table:institute_breakup_survey}
    \begin{tabular}{@{}lc@{}}
        \toprule
        \textbf{Institute} & \textbf{Number of Users} \\
        \midrule
        IIITD & 16 \\
        BITS & 21 \\
        SVCET & 5 \\
        BLENDNET & 3 \\
        \bottomrule
    \end{tabular}
\end{table}

In addition to the survey, we conducted 25 personalized, semi-structured interviews \cite{adeoye2021research} with participants who had completed at least five coding questions on the Sakshm AI  platform. The participant demographics are summarized in Table  (see Table \ref{table:demographic}), which includes 15 females and 10 males. Interviewees were selected from the survey pool to ensure the representation of both users who had interacted with Disha and those who had not.

These interviews aimed to gain deeper insights into individual user experiences, challenges, and overall perceptions of the platform and its AI assistant, Disha. The sessions generated 424.16 minutes of recorded content, which were analyzed using thematic analysis \cite{bingham2021deductive} to identify recurring patterns and themes in user feedback. This qualitative data offered a comprehensive understanding of participants’ experiences, uncovering nuanced insights into their interactions with the platform.

All interviews were transcribed for detailed analysis. The interview questions can be found in Appendix A.

\begin{table}[t]
    \centering
    \small  
    \caption{\textbf{Interview Participants Demographics }}
    \label{table:demographic}
    \begin{tabular}{@{}ll@{}} 
        \toprule
        \textbf{Category} & \textbf{Details} \\
        \midrule
        \textbf{Gender} & Female: 10, Male: 15 \\
        \textbf{Age} & Min: 19, Max: 31, Avg: 21.80, SD: 2.66 \\
        \textbf{Occupation} & Student: 21, Phd Scholar : 3, Working Professionals: 1  \\
        \bottomrule
    \end{tabular}
\end{table}

\subsection{Data Analysis}

For the analysis of Sakshm AI, a quantitative survey with 46 participants was taken, capturing broad trends in user engagement, satisfaction, and challenges faced while using the platform. The survey focused on various aspects of the Sakshm AI platform, including its features, effectiveness, and the role of AI in supporting learning \cite{chiu2024teacher}. The survey data provided valuable insights into user experiences, offering a foundation for further qualitative exploration.

For the qualitative analysis \cite{ezzy2013qualitative} of Sakshm AI, we conducted a thematic analysis of interview transcripts \cite{bingham2021deductive} and survey responses collected from 25 participants (10 females and 15 males) representing diverse demographics in terms of age, educational background, and coding experience\cite{mhasakar2024comuniqa}. These participants actively used the Sakshm AI platform, attempted coding problems, and interacted with the AI assistant, Disha, for guidance. The interviews, totaling  424.16 minutes of recordings, provided a rich dataset to explore user experiences and perceptions in depth.

Using an inductive approach \cite{proudfoot2023inductive,bingham2021deductive}, we analyzed recurring themes and patterns related to their engagement with the platform, problem-solving strategies, and the perceived impact of AI-based tools on learning. The themes identified included user perceptions of platform features, the role of AI assistance in learning and problem-solving, challenges faced while using the platform, and recommendations for improving both the Sakshm AI platform and Disha. Specifically, themes such as the effectiveness of step-by-step guidance versus direct answers, the confidence in implementing concepts post-interaction, and comparisons between AI-based and human assistance were explored in detail. This thematic analysis provided nuanced insights into how users engaged with Sakshm AI, highlighting its strengths and areas for refinement while complementing the quantitative trends observed in the survey data.

\subsection{Limitations}
Currently, Sakshm AI operates as an independent learning platform and is not integrated into any formal teaching curriculum, limiting its direct applicability within institutional, educational frameworks. The platform exclusively supports coding in four programming languages and remains in the developmental stage, restricting its versatility and scalability for a broader range of subjects and languages. Additionally, the Disha chatbot is only activated during Data Structures and Algorithms (DSA) problem-solving sessions, preventing users from accessing it to revise doubts, engage in mock interviews, or utilize forthcoming elaborative features yet to be implemented. This research is based on a seven-month study period from July to January, precluding the ability to draw long-term conclusions about Sakshm AI’s efficacy and impact. Furthermore, the current dataset is predominantly sourced from top-tier engineering colleges \cite{budhiraja2024s}, which may not fully represent the diverse experiences and challenges students from tier 3 institutions face. Future studies aim to incorporate data from various educational institutions to address this imbalance and provide a more comprehensive understanding of Sakshm AI’s effectiveness across different academic environments\cite{joshi2023let}. These limitations highlight the need for ongoing development and broader data collection to improve the functionality and generalizability of the platform.

\section{Findings}\label{sec:evaluation}
\subsection{ Quantitative analysis }
The quantitative analysis explores how users interacted with the Sakshm AI platform, particularly the impact of difficulty level, solving time, chat usage, hourly trends, and user-level performance. This section draws insights from multiple data dimensions to highlight user behavior and the effectiveness of the platform’s features.
Our quantitative analysis is based on the user logs collected from the platform between 2024-07-18 13:31:28 and 2025-01-23	13:58:24. The user-logs had the following properties:-
\begin{table*}[h]
    \centering
    \begin{tabular}{|c|c|}
        \hline
        \textbf{Difficulty levels of questions} & basic, easy, medium, hard \\ \hline
        \textbf{Number of unique users} & 1171 \\ \hline
        \textbf{Colleges} & None(957 users), BITS(35), TestUsers(8), IIITD(75), UPES(34), SVCET(61) \\ \hline
        \textbf{Total attempts} & 4109 ( 3471 Closed, 638 not closed) \\ \hline
        \textbf{Number of attempts v.s. chat} & 1038/4109 initiated a chat \\ \hline
    \end{tabular}
    \caption{Properties of the user-logs}
    \label{tab:label_placeholder}
\end{table*}

\subsubsection{\textbf{Impact of Difficulty Level on User Behavior}}\hfill

\textbf{Basic:} There are 34 unique questions, attempted 2230 times, making it the most heavily attempted category. This suggests that users are more comfortable starting with basic problems. A total of 661 chat instances were initiated, corresponding to 1734 messages. This indicates that the users engaged in an average of .77 chat messages per attempt for basic questions.  

\textbf{Easy:} 90 unique questions were attempted 1084 times, showing an intermediate level of engagement. Here, there were 245 instances of chat, with 760 messages exchanged. Here also, the users interacted with chat messages quite a bit i.e., 0.71 chat messages per attempt

\textbf{Medium:} With 68 unique questions attempted 641 times, this difficulty level attracts moderate attempts compared to basic and straightforward. Here, there were 112 instances of chat, generating 519 messages. This maximizes user interaction with the chats, having .81 messages per attempt. This indicates that users might most use the chat feature for medium-difficulty questions. 

\textbf{Hard:} 16 unique questions were attempted only 154 times, showing a significant drop in user engagement, likely due to increased complexity. There were a total of 21 instances of chat, with only 80 messages exchanged. This has the least chat interaction, with only 0.5 messages per attempt. This shows that users might seek the Socratic chatbot's help for medium, easy, and basic questions and less for hard ones. This data is highlighted in Table: \ref{tab:usage_diff}. 

Also, Table: \ref{tab:comp_diff} shows that the completion rate of all four difficulty levels is very similar, but for beginner, easy, and medium difficulty problems, around 11-12\% completion happened with chat support, whereas only 5\% completion happened with chat support for hard problems. This further highlights that the Socratic AI chatbot is useless for complex problems. 
\begin{table*}[h]
    \centering
    \begin{tabular}{|c|c|c|c|}
        \hline
        \textbf{Difficulty} & \textbf{Number of unique questions} &\textbf{Total number of attempts} &\textbf{Total number of chat messages} \\ \hline
        basic & 34 & 2230 & 1734\\ \hline
        easy & 90 & 1084 & 760 \\ \hline
        medium & 68 & 641 & 519 \\ \hline
        hard & 16 & 154 & 80\\ \hline
    \end{tabular}
    \caption{Chat usage with problem difficulty}
    \label{tab:usage_diff}
\end{table*}

\begin{table*}[h]
    \centering
    \begin{tabular}{|c|c|c|c|}
        \hline
        \textbf{Difficulty} &\textbf{Total number of attempts} & \textbf{Total number of attempts closed} &\textbf{Total number of attempts closed with chat} \\ \hline
        basic & 2230 & 1767 & 199\\ \hline
        easy & 1084 & 961 & 122\\ \hline
        medium & 641 & 603 & 72\\ \hline
        hard & 154 & 138 & 6\\ \hline
    \end{tabular}
    \caption{Completion rate with problem difficulty and chat usage}
    \label{tab:comp_diff}
\end{table*}

\subsubsection{\textbf{Hourly Activity Trends}}
\begin{figure*}[h]
   \centering
    \includegraphics[width=0.8\textwidth]{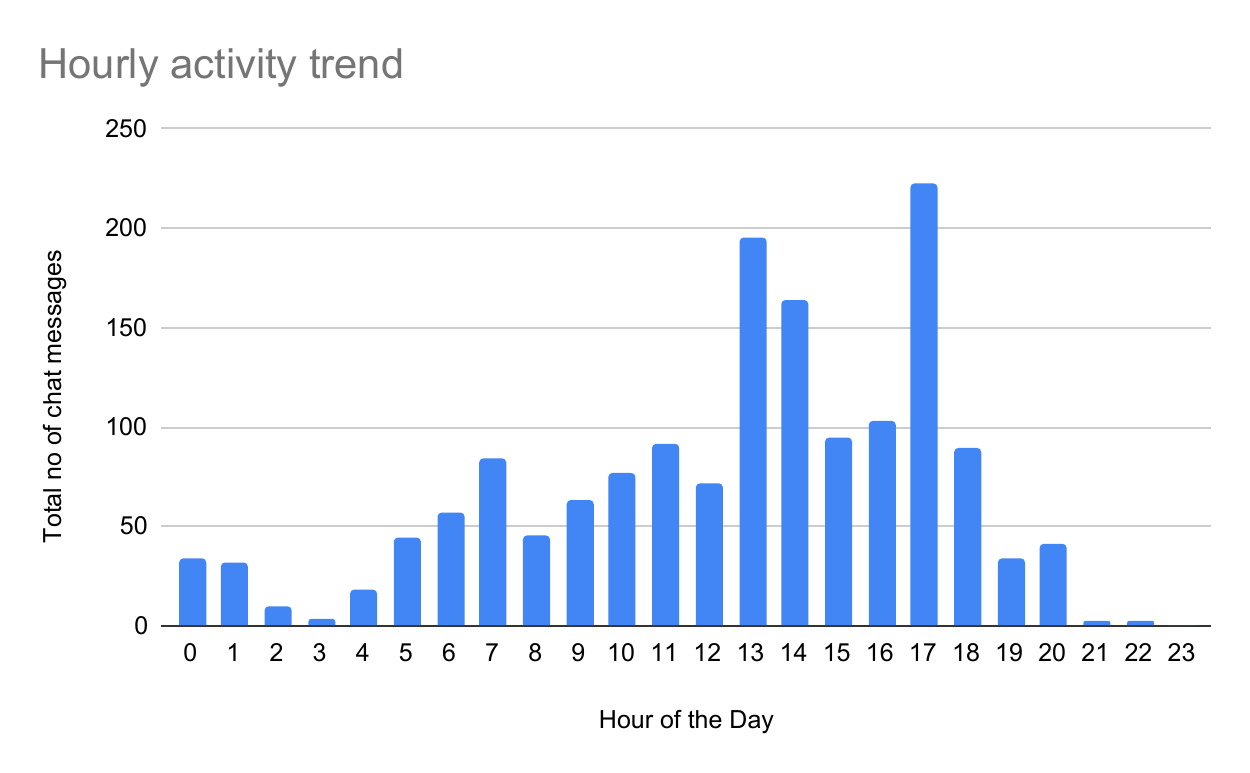}  
    \caption{Hourly Activity Trend of users}
    \label{fig:excel_graph}
\end{figure*}

\hfill
Chat activity on the Sakshm AI platform is minimal during late-night hours (12 AM - 5 AM), with the lowest engagement at 3 AM (4 messages). Activity gradually rises from 6 AM, peaking at 7 AM (84 messages). A significant increase occurs in the afternoon, with a sharp spike at 1 PM (195 messages) and the highest engagement at 5 PM (223 messages). After 6 PM, chat activity declines, dropping nearly to zero by 9 PM. This pattern suggests structured engagement, with peak afternoon and early evening activity aligning with post-school or study sessions. This is shown in Fig: \ref{fig:excel_graph}. 

\subsubsection{\textbf{College-Wise Analysis}}\hfill

\begin{figure*}[h]
   \centering
    \includegraphics[width=0.8\textwidth]{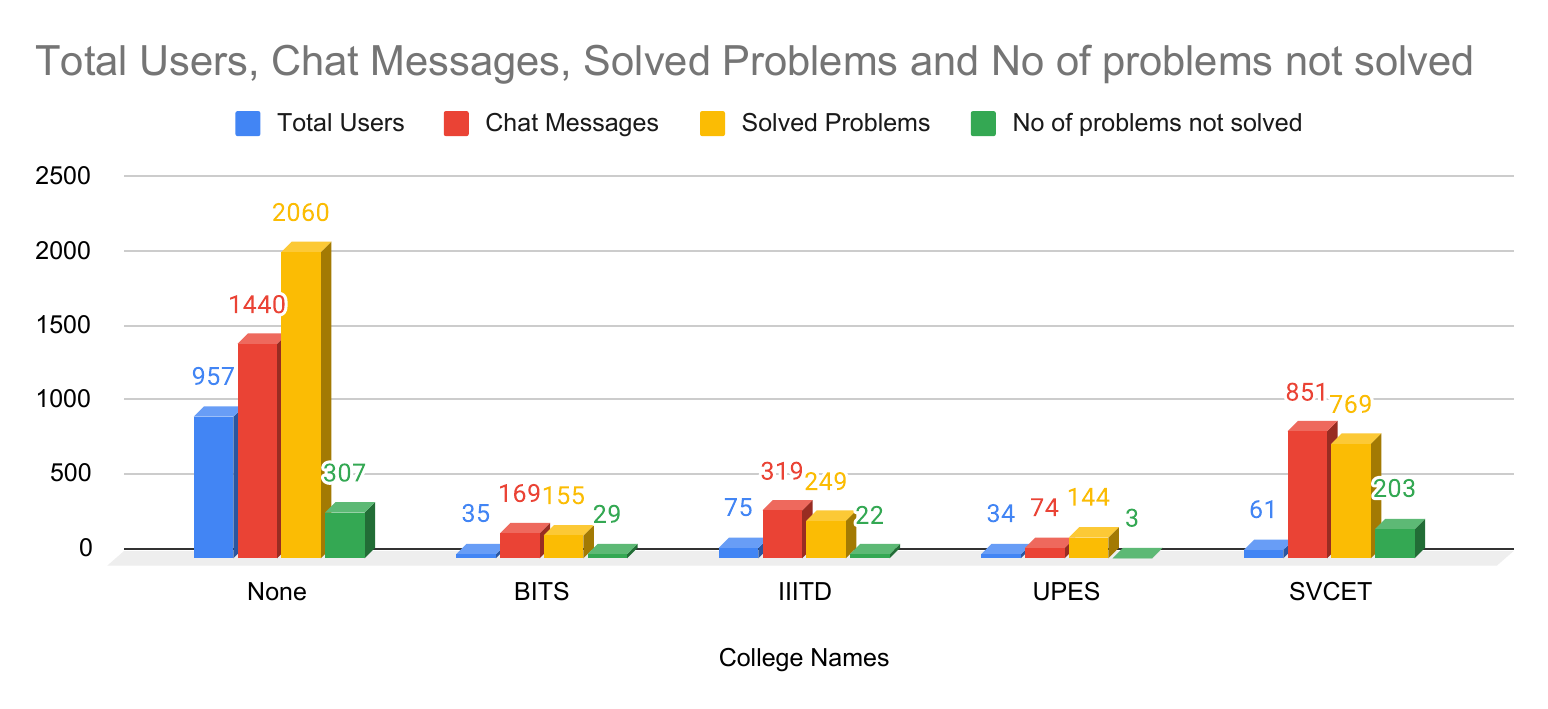}  
    \caption{College-Wise Analysis}
    \label{fig:college_graph}
\end{figure*}

The Sakshm AI platform analysis shows that the largest group of users (957) did not specify a college, contributing the highest chat messages (1440) and solved problems (2060), with a moderate number of unsolved problems (307). Among specific colleges, IIITD had the highest problem-solving efficiency (249 solved, 22 unsolved), while BITS (155 solved, 29 unsolved) and UPES (144 solved, only 3 unsolved) also showed strong performance. SVCET users engaged the most in chat (851 messages) but struggled with the highest failure rate (203 unsolved problems out of 769 attempts), suggesting more reliance on support. This is being referred to in Fig: \ref{fig:college_graph}.

\subsubsection{\textbf{User-Level Performance}}\hfill

The analysis categorizes users into four quartiles based on the number of problems attempted, providing insights into engagement patterns, problem-solving success, and chat usage. The first quartile (Q1) comprises the least engaged users (518 users), who attempted only one problem on average, closing 73.5\% of them with minimal chat interactions (4.6\%). The second quartile (Q2) includes moderately engaged users (323 users), attempting two problems on average, with an improved success rate of 90\% and slightly higher chat reliance (7.1\%). The third quartile (Q3) represents highly engaged users (68 users), who attempted three problems on average, with a strong success rate (91\%) and a significant increase in chat usage (30.8\%). The fourth quartile (Q4) consists of super users (261 users), attempting an average of 10.5 problems, with an 85\% success rate and moderate chat reliance (12.7\%).

\textbf{Key insights:} Most users fall in Q1 and Q2, indicating that a significant proportion of users have low engagement. Higher engagement levels (Q3 and Q4) correspond to increased problem-solving efficiency and greater use of chat support. Notably, Q3 users show the highest chat reliance, suggesting that users seeking assistance perform better. Super users (Q4) are strongly committed to problem-solving, attempting the most problems with effective but controlled chat support. These findings highlight the need for targeted interventions to improve engagement in lower quartiles and optimize chat-based assistance for active users.
Table : \ref{tab:quartile_analysis} shows the user data :-


\begin{table*}[h]
    \centering
    \small 
    \setlength{\tabcolsep}{4pt} 
    \begin{tabular}{lcccccc}
        \toprule
        \textbf{Quartiles} & \textbf{Threshold} & \textbf{Users} & \textbf{Avg Attempts} & \textbf{Avg Closed} & \textbf{Avg Closed (Chat)} & \textbf{Avg Closed (No Chat)} \\
        \midrule
        Q1 (Least Engaged)  & 1   & 518  & 1.00  & 0.736  & 0.046  & 0.689 \\
        Q2 (Moderate Users) & 2   & 323  & 2.00  & 1.802  & 0.071  & 1.731 \\
        Q3 (Highly Engaged) & 3   & 68   & 3.00  & 2.721  & 0.309  & 2.412 \\
        Q4 (Super Users)    & 165 & 261  & 10.50 & 8.900  & 1.272  & 7.628 \\
        \bottomrule
    \end{tabular}
    \caption{User engagement analysis based on quartiles}
    \label{tab:quartile_analysis}
\end{table*}

\subsection{Qualitative Analysis}

Our findings show that students preferred the Socratic AI chatbot, Disha, for its ability to foster critical thinking and problem-solving, making it a valuable alternative to platforms that provide direct solutions. Sakshm AI stood out for its structured learning experience, topic-wise practice, and user-friendly features, though some users highlighted challenges such as technical issues and limited question variety. The platform’s feedback and reporting system were highly appreciated for offering actionable insights on code quality and efficiency and promoting professional coding habits. While users valued the accessibility of AI tools like Disha, they also emphasized the importance of balancing AI assistance with human guidance for deeper learning. Compared to other tools, Sakshm AI’s integration of contextual hints, company-specific questions, and engagement features positioned it as a comprehensive platform for coding education.

We conducted a thematic analysis of 25 semi-structured interviews, totaling 424.16 minutes of recordings, applying open, axial, and selective coding  (see Section~\ref{sec:codebook}). We identified key themes related to  AI-assisted learning and problem-solving strategies through open coding. Axial coding established relationships between themes, such as the trade-off between AI guidance and independent learning. Finally, selective coding synthesized these insights into core themes, highlighting user perceptions, AI-human comparisons, and platform usability. 

We begin by understanding the participants' platforms for study and coding practice (Section~\ref{sec:study-coding}) and doubt resolution (Section~\ref{sec:doubt-resolution}). Then, we discuss the effectiveness of the platform (Section~\ref{sec:effectiveness}) and the challenges in using it (Section~\ref{sec:challenges}). We then separately discuss the platform's feedback and reporting feature (Section~\ref{sec:feedback-reporting}) and its AI chatbot, Disha, which provides Socratic guidance (Section~\ref{sec:socratic-guidance}), the need to balance AI and human assistance to maximize student learning (Section~\ref{sec:ai-human-balance}), and finally, a short comparison of the Sakshm AI platform with other tools in the market (Section~\ref{sec:comparison-tools}).

\subsubsection{\textbf{Study and Coding Practices}}`\hfill\\
\label{sec:study-coding}

We interviewed participants regarding the platforms and resources they use for coding practice and what they like about each platform. The participants discussed their preferences for coding platforms to practice and enhance their skills. Popular choices included platforms like LeetCode\cite{cozzolino2021digital,coignion2024performance,sahay2020supporting}, HackerRank\cite{orts2018analysis,vamsi2020classifying}, CodeChef\cite{malik2022study,mirzayanov2020codeforces}, and GeeksforGeeks\cite{malik2022study}. These platforms were appreciated for their extensive repositories of coding problems, gamification features, and user-friendly interfaces. Some participants preferred LeetCode and HackerRank for coding practice, while others primarily used CodeChef and Sakshm AI. One participant shared, \textbf{\emph{"ChatGPT is more effective, especially in terms of coding help and debugging."}} - (P3), highlighting its usefulness in solving problems, providing potential outputs, and identifying errors. When asked about the resources they use for learning, many participants mentioned frequently using AI-based tools like ChatGPT for debugging, learning, or understanding specific concepts\cite{coello2024effectiveness,biswas2023role,tian2023chatgpt}. Coursera, Udemy, and other structured course platforms were also mentioned for their comprehensive learning modules. A respondent shared, \textbf{\emph{"I mostly use ChatGPT to clarify doubts or write some code. I haven’t tried other platforms."}} - (P4). \textbf{\emph{"I rely on platforms like Coursera and Udemy for structured courses."}} - (P5). These insights reveal how learners integrate AI tools and traditional online courses to enhance their coding skills.

We asked the interviewees about their study and coding habits and the specific features they valued in learning and coding platforms. Participants mentioned that their motivation to code stemmed from gamification features, competitive programming, and long-term career aspirations. Platforms like LeetCode helped maintain consistency through streak features and achievement badges. One participant shared, \textbf{\emph{"I use LeetCode because it has this streak feature that motivates me to maintain a 30-day streak and earn a badge."}} - (P1). \textbf{\emph{"I participate in weekly coding contests, which I find engaging due to the group dynamics and competition."}} - (P7). This highlights the importance of motivation and structured challenges in sustaining coding practice.Some participants preferred visual learning tools such as video tutorials and interactive diagrams. Platforms like YouTube were favored for explaining complex concepts and algorithms in an engaging visual format. \textbf{\emph{"I use YouTube a lot for tutorials because seeing someone explain a concept visually makes it easier to understand."}} - (P6). \textbf{\emph{"YouTube videos work best for me when I need to visualize a concept, especially algorithms or DSA patterns."}} - (P7).

\subsubsection{\textbf{Resolving Doubts and Seeking Help}}\hfill\\
\label{sec:doubt-resolution}

Participants were asked about the resources they rely on for doubt resolution. In response, many mentioned frequently using online resources and search engines to resolve coding doubts. Popular platforms such as Stack Overflow, W3Schools, LeetCode, and GeeksforGeeks were considered reliable sources for theoretical explanations and practical solutions. One participant explained, \textbf{\emph{"I usually start by Googling it. There’s almost always someone else who’s had the same question, so Stack Overflow is a big help for coding."}} - (P6). \textbf{\emph{"I use GeeksforGeeks and W3Schools for theoretical knowledge. ChatGPT helps with straightforward answers but sometimes provides wrong ones, so I cross-check with verified platforms."}} - (P8). Twenty out of twenty-five participants mentioned using AI-based tools, particularly ChatGPT, to resolve doubts. Participants appreciated the instant and concise responses provided by AI, especially for debugging, code suggestions, and explaining complex logic. Some participants shared, \textbf{\emph{"I use ChatGPT in difficult situations."}} - (P9), and \textbf{\emph{"I ask ChatGPT to explain it step-by-step."}} - (P10).

Despite the increasing reliance on online and AI-based tools, four participants still valued human support for resolving doubts, particularly for personalized and context-specific explanations. Professors, peers, and lab instructors played a significant role. One participant noted, \textbf{\emph{"If the professor is available, I prefer consulting them first. Otherwise, I try to solve the issue myself or use AI tools like ChatGPT."}} - (P4). \textbf{\emph{"Sometimes classmates are the best resource for resolving doubts, especially for shared coursework."}} - (P6).
In the context of doubt resolution on the Sakshm AI platform, participants highlighted the unique role of Sakshm AI and its Disha chatbot in fostering a problem-solving mindset. Unlike traditional AI tools, Disha’s hints and Socratic approach encourage users to think critically rather than providing direct answers. One participant shared, \textbf{\emph{"Disha is helpful because it encourages problem-solving by providing hints rather than direct answers."}} - (P12). \textbf{\emph{"I like that Disha doesn’t spoon-feed answers. Instead, it gives me hints, making me work through the solution independently."}} - (P11). This approach not only aids learning but also builds confidence in tackling complex coding challenges.

\subsubsection{\textbf{Effectiveness of Sakshm AI Platform}}\hfill\\ 
\label{sec:effectiveness}

We asked participants specific questions about the Sakshm AI platform. The group included a mix of students and working professionals who had used the Disha AI chatbot. When asked about the platform’s user interface, most participants appreciated its simplicity and organization, which made it user-friendly and easy to navigate. Features like dark mode and the structured interface were particularly well-received. One participant shared, \textbf{\emph{"I got familiar with the platform in just a few minutes... navigating through multiple options was very intuitive."}} - (P8). However, a few participants were less enthusiastic, noting that its \textbf{\emph{"questions focus on data structures,"}} which they were still learning. Participants also highlighted features like topic-wise practice, difficulty levels, and test cases, which they felt contributed to thorough learning. One participant explained, \textbf{\emph{"The number of test cases is impressive; questions have over ten test cases, covering edge cases and ensuring my code is efficient."}} - (P8).

We also asked participants about the platform’s topic coverage, particularly regarding company interview preparation. Fifteen students appreciated the platform’s extensive content and alignment with industry standards. The platform's repository included curated questions from companies like Flipkart, Amazon, and Google and time complexity analysis to improve code efficiency. \textbf{\emph{"Sakshm AI provides previously asked questions from companies like Flipkart, Accenture, Amazon, and Google."}} - (P12). \textbf{\emph{"It also gave me ideas on how to optimize my code... whenever my code complexity was O(n²), it suggested an O(n log n) approach."}} - (P13). We then asked participants how they felt about having an AI assistant on the platform. Seventeen out of twenty-five participants viewed it as a standout feature of the Sakshm AI platform. They valued the chatbot’s ability to guide problem-solving without directly providing answers. One participant remarked, \textbf{\emph{"The chatbot is unique because it remembers my code context. This saved me the time and effort of switching tabs and re-entering contexts that I generally do with ChatGPT."}} - (P10). However, one participant mentioned, \textbf{\emph{"I could not figure out that it was an AI tutor as it just looked like a generic chatbot."}} - (P14).

\subsubsection{\textbf{Challenges in Using the Platform}}\hfill\\
\label{sec:challenges}

We asked participants about aspects of the platform they did not like. Six participants expressed that the platform could include more varied and real-world-oriented exercises. One participant remarked, \textbf{\emph{"There could be more variety in the exercises, and it should have more questions."}} - (P6). \textbf{\emph{"The problems are pretty basic... the problems nowadays are more twisted, like in online tests."}} - (P15). The need for dynamic, multi-concept challenges emerged as a recurring theme. Four participants mentioned technical issues such as slow performance, bugs, and inconsistencies in the user interface. These issues occasionally disrupted their workflow and caused frustration. For example, one user shared, \textbf{\emph{"The platform takes a long time to show results after submitting code, sometimes even hours."}} - (P16). \textbf{\emph{"The platform froze while I was coding, and when I refreshed, all my progress was lost. That was very frustrating."}} - (P4).

When asked about the features of the AI chatbot that they didn’t like, five participants found its responses occasionally repetitive, vague, or overly focused on certain aspects. One participant mentioned, \textbf{\emph{"Sometimes the responses are too repetitive and long, making it difficult to understand."}} - (P3). \textbf{\emph{"Disha focuses too much on syntactical errors and doesn't point out logical issues until the syntax is perfect."}} - (P8). The platform's interface and user engagement features were another area where participants saw potential for improvement. Suggestions included adding more community-based features, intuitive design updates, and gamification elements. One participant observed, \textbf{\emph{"The platform lacks community-based interaction features like leaderboards or forums."}} - (P15). \textbf{\emph{"Having more interactive, gamified features could improve engagement."}} - (P3). We also asked participants about the Sakshm AI platform's feedback. Two of them mentioned that the platform should provide more actionable feedback and deeper analysis of their performance during assessments. They suggested enhancements like explanations for wrong answers, efficiency comparisons (e.g., comparing the efficiencies of different approaches to solving a question), and categorized feedback. They also suggested including the problem-solving timeline in the final review. \textbf{\emph{"Reviewing problem-solving timelines would benefit long-term improvement."}} - (P15).

\subsubsection{\textbf{Feedback and Reporting}}\hfill\\
\label{sec:feedback-reporting}

The participants were asked about the feedback report generated by the platform after each submission. Regarding the depth and detail of the feedback, ten participants appreciated the reports for their focus on code quality, efficiency, and readability. However, many felt they lacked sufficient detail in explaining why a solution worked or failed. One participant mentioned, \textbf{\emph{"The feedback report was helpful, and I specifically liked the code quality review as it can help us write industry-level code."}} - (P5), while another expressed, \textbf{\emph{"I wish the feedback report had more depth on why a solution worked or didn’t work."}} - (P16).Sakshm AI’s grading system was generally praised for encouraging incremental learning, but three users found the scoring criteria inconsistent or overly focused on certain aspects like comments or formatting. One user stated, \textbf{\emph{"Disha’s grading is excellent, providing knowledge incrementally instead of just direct answers."}} - (P9). On the other hand, another participant shared, \textbf{\emph{"It would always prompt me to add comments for better code cleanliness, but I never scored a perfect 100."}} - (P1).

While the feedback system was effective for beginners and intermediate learners, participants noted a gap in resources and guidance for advanced problem-solving. One participant commented, \textbf{\emph{"Sakshm AI should provide more resources to help bridge the gap between beginner and advanced learners."}} - (P16). \textbf{\emph{"Adding more examples or asking follow-up questions would help bridge the gap between basic and advanced learners."}} - (P5).This suggests a demand for more advanced coding challenges, time-based competitive scenarios, and pathways to transition learners to industry-ready standards.

\subsubsection{\textbf{Impact of Socratic Guidance}}\hfill\\
\label{sec:socratic-guidance}

Participants were asked about the Socratic method used by Sakshm AI’s AI chatbot, Disha. Fifteen participants appreciated this approach, as it fosters independent thinking and critical problem-solving skills. Many users noted that the hint-based guidance encouraged them to approach problems methodically. One participant shared, \textbf{\emph{"Disha doesn’t give direct solutions; it provides hints that help me think and solve problems myself, maintaining my integrity in learning."}} - (P18). \textbf{\emph{"The one that gives hints is better because it encourages us to think and doesn’t stop the thought process."}} - (P19). Four participants mentioned that the usability of the Socratic method depends on the complexity or urgency of the problem. For instance, one participant remarked, \textbf{\emph{"Step-by-step guidance is better for complex problems because it helps me understand the process, whereas direct answers are fine for smaller questions or syntax issues."}} - (P6). \textbf{\emph{"For learning, I’d prefer hints. But during busy times, direct answers are practical."}} - (P20).

Some participants utilized the Socratic approach to further improve their code. One participant explained, \textbf{\emph{"It asked me to think differently. For example, it identified that my approach was traversing the list twice instead of once and encouraged me to reduce it."}} - (P7). \textbf{\emph{"I once tried an inefficient solution, and Disha pointed out a better approach to the problem. It was insightful."}} - (P8). Such feedback enhanced learning and instilled confidence in tackling future problems. When asked about the limitations of the Socratic method, nine participants expressed dissatisfaction, particularly in time-sensitive situations or with more challenging problems. One user mentioned, \textbf{\emph{"Hints can be less helpful in time-sensitive situations."}} - (P12). \textbf{\emph{"For hard questions, hints aren’t always enough, and sometimes I end up switching to ChatGPT for a full solution."}} - (P2). These responses illustrate the approach’s limitations when users need quick resolutions or face complex problems that require direct guidance.

\subsubsection{\textbf{Balancing AI and Human Assistance}}\hfill\\
\label{sec:ai-human-balance}

We asked the participants a few questions comparing traditional human assistance in teaching versus AI assistance. When asked which of the two they feel is more accessible and which is more accurate, twelve appreciated the round-the-clock availability and speed of AI tools. Yet, seven of them highlighted the irreplaceable value of human tutors in providing personalized insights. One participant noted, \textbf{\emph{"AI is fast and accessible anytime, which is great, but human tutors understand my questions better and can adapt their answers based on my level of understanding."}} (P6). \textbf{\emph{"AI tools are accessible 24/7, but they lack the deeper insights that a professor or mentor can provide."}} (P7). While AI tools like Disha and ChatGPT are highly convenient, users acknowledged that human instructors bring creativity and adaptability that enhance foundational learning. Some participants also expressed concerns about over-reliance on AI tools, with nine fearing it might hinder independent thinking. One user remarked, \textbf{\emph{"AI is helpful, but it can reduce our thinking power. We often rely on it rather than trying to solve the problem independently."}} (P12). \textbf{\emph{"I remind myself to practice solving problems independently, too, even when AI tools provide quick answers."}} (P5). This balance between leveraging AI for efficiency and maintaining critical thinking emerged as a key consideration for effective learning.

We asked the participants how they envisioned the perfect integration of AI tutors in education. One participant responded, \textbf{\emph{"For a successful AI integration in education, I’d envision a system that gives hints, evaluates a student’s progress, and then offers snippets or sub-solutions when needed like an in-person tutor would."}} (P21). Many participants suggested a balanced approach combining AI and human guidance would be ideal. While AI tools are helpful for hints and repetitive tasks, human tutors add value in complex scenarios requiring adaptability. One participant explained, \textbf{\emph{"I would still consult faculty for in-depth understanding if they know the topic. But when faculty are unavailable, AI is helpful."}} (P13). This indicates that users see the potential for AI to complement rather than replace human assistance. Participants also perceived the trade-off between the speed offered by AI tools and the depth of understanding provided by human tutors. One participant explained, \textbf{\emph{"AI tools make the learning process very efficient, but over-reliance can reduce conceptual understanding."}} (P4). Another observed, \textbf{\emph{"If I need to finish work quickly, I’d prefer a direct answer. But when learning, I think AI tools like Disha that give hints are better for long-term understanding."}} (P1). These perspectives underline the importance of using AI tools judiciously to balance immediate productivity with deeper learning.

\subsubsection{\textbf{Comparison with Other Tools}}\hfill\\ 
\label{sec:comparison-tools}

Most participants were exposed to other coding platforms, allowing them to compare the Sakshm AI platform with them. Fifteen participants expressed that Sakshm AI’s AI chatbot, Disha, stood out for promoting critical thinking by providing hints and guidance rather than direct answers. Participants frequently contrasted Disha with platforms like ChatGPT, highlighting that Disha fosters logical problem-solving: \textbf{\emph{"Unlike ChatGPT, where you can paste code and get answers, Disha makes you think critically and approach the problem logically."}} (P2). \textbf{\emph{"Disha forces you to think critically about the problem and solution, unlike platforms where you can just copy solutions."}} (P22). This focus on engagement and understanding makes Sakshm AI more effective for learning, especially for students aiming to build problem-solving skills. Seven participants appreciated that Sakshm AI integrates features that enhance structured learning and reduce the need to switch between tools. The platform’s ability to maintain context during problem-solving was also highly valued: \textbf{\emph{"Disha maintained the context of my code, my last query, and gave contextual guidance."}} (P23). Additionally, features like company-specific questions, free complexity analysis, and integrated hints were highlighted as unique. Another participant shared: \textbf{\emph{"LeetCode offers complexity analysis but only for a limited number of free questions per day, whereas Sakshm AI provides this feature for free."}} (P3). This integration ensures a seamless and comprehensive learning experience. 

When asked about specific aspects of the platform’s UI, five participants appreciated the user-friendly design of Sakshm AI but pointed out areas for improvement in usability and additional tools. For example, one participant mentioned: \textbf{\emph{"The interface needs an autosave feature. I lost my code’s progress after the page refreshed."}} (P25). Others suggested adding career-oriented features such as a resume builder and better dark mode functionality. A user remarked: \textbf{\emph{"Adding features like a resume builder or interview prep tools would make the platform more complete."}} (P8). Despite these areas for enhancement, the platform’s intuitive layout and focus on structured learning were widely appreciated. Participants compared Sakshm AI with platforms like LeetCode, Stack Overflow, and GitHub Copilot, valuing its structured learning approach and unique engagement features. One participant noted: \textbf{\emph{"Sakshm AI is less overwhelming than other platforms, especially when practicing under time constraints."}} (P15). Another stated: \textbf{\emph{"Sakshm AI’s hint-based approach and detailed feedback reports stand out compared to other AI coding platforms."}} (P22). These features position Sakshm AI as a well-rounded platform, especially for students preparing for technical interviews or working on foundational coding skills.

\section{Discussion}\label{sec:discussion}
The discussion interprets the findings of our study, addressing the research questions and situating Sakshm AI  within the broader context of AI-driven educational tools. By exploring how Disha’s Socratic approach enhances critical thinking, identifying the features that make Sakshm AI effective for structured learning, and examining the balance between AI and human assistance, this section provides insights into the platform's contributions and limitations. Comparisons with existing tools such as ChatGPT, LeetCode, and CodeAid reaffirm Sakshm AI’s strengths while highlighting its unique features. The section also presents practical recommendations for stakeholders and outlines avenues for future research to further advance the impact of AI in education.

\subsection{\textbf{Sakshm AI's Socratic chatbot enhances critical thinking}}

The Socratic approach employed by Disha encourages critical thinking and problem-solving by guiding users through hints rather than providing direct answers. This aligns with existing research on pedagogical AI tools like CodeAid and Cipherbot, which emphasize the importance of engaging users in the learning process rather than spoon-feeding solutions. Similarly to the research findings of tools like CodeHelp \cite{liffiton2023codehelp}, Sakshm AI users reported that Disha fosters logical thinking by prompting students to consider alternative approaches.
However, unlike platforms such as ChatGPT, which often provide direct answers, Disha’s approach intentionally delays gratification, allowing users to better understand coding problems. As one participant noted, \textit{"Unlike ChatGPT, Disha makes you think critically and approach the problem logically."} This highlights a novel aspect of Sakshm AI, where the balance between assistance and independent thought is carefully calibrated to promote learning.
Although students use many other platforms like LeetCode extensively for coding practice, Sakshm AI sets itself apart by integrating AI assistance through its Disha chatbot, offering real-time hints and detailed feedback that other platforms lack. A user pointed out, \textit{“I use HackerRank and LeetCode, but there’s no AI assistance there like in Sakshm AI,” }emphasizing the added value of AI in enhancing the learning experience. This AI integration not only aids in understanding problem-solving techniques but also provides comprehensive code analysis, offering a more interactive and supportive environment for learners than traditional platforms like LeetCode.
Many students mentioned using ChatGPT for doubt resolution, debugging, and code analysis, but this requires them to paste their entire code in the prompt. Also, ChatGPT doesn't have the context of how a student went about writing that code, like whether they stopped too much while writing a particular line of code or whether they arrived at it after multiple iterations, etc. Also, to enforce Socratic Guardrail on it explicitly, and it is very easy to break it. In contrast, Sakshm AI maintains student context and offers a more integrated and interactive learning environment. For example, \textit{“Unlike ChatGPT, where you can paste code and get answers, Disha makes you think critically and approach the problem logically,”}. Also, Disha is coded to give only Socratic Guidance, and that too, is what is relevant to the question at hand. So, if the student tries to ask for direct answers or starts asking irrelevant questions, it brings the conversation back to the specific question. This approach not only supports a more focused learning process but also promotes a deeper understanding of coding concepts, setting Sakshm AI apart from tools like ChatGPT and Gemini. 

\subsection{\textbf{Impact of Sakshm AI's feedback on learning}}

Sakshm AI integrates features such as contextual hints, feedback reports, and topic-wise practice, which contribute to a seamless learning experience. These findings resonate with existing platforms like LeetCode and CodeAid, which focus on structured learning and complexity analysis. However, Sakshm AI goes a step further by offering company-specific questions and detailed feedback on code quality, including naming conventions and optimization suggestions.
In contrast to Codehelp \cite{liffiton2023codehelp}, which lacks comprehensive feedback mechanisms, Sakshm AI’s ability to evaluate multiple dimensions of code—such as readability, efficiency, and adherence to professional standards emerges as a key differentiator. For instance, \textit{"LeetCode offers complexity analysis but only for a limited number of free questions per day, whereas Sakshm AI provides this feature for free."} Such offerings enhance Sakshm AI’s value as a platform that bridges the gap between beginner and advanced coding learners.

\textbf{How can Sakshm AI balance AI assistance and human tutoring to optimize student learning outcomes?}
Balancing AI and human assistance is crucial, as users emphasized the importance of both accessibility and personalized guidance. Research on Cipherbot and Codehelp \cite{liffiton2023codehelp} underscores the need for AI tools to complement rather than replace human tutors. Sakshm AI aligns with this philosophy by providing scalable support through its 24/7 chatbot while still recognizing the value of human mentoring for foundational learning and complex scenarios.
However, Sakshm AI distinguishes itself through its focus on fostering independent problem-solving, as one user noted, \textit{"AI tools are great for hints, but the first step in solving should be your own thinking."} This suggests that platforms like Sakshm AI can act as a bridge, allowing students to transition from AI-guided learning to self-reliance. Many students prefer using Disha when they have the time to engage thoroughly with the material, appreciating how it guides them without offering direct answers. However, when faced with particularly challenging problems or under time constraints, students often turn to ChatGPT for more comprehensive and immediate solutions. For example, some students have mentioned that if they need to finish work quickly, they prefer the direct answers provided by ChatGPT, whereas Disha is favored for its ability to foster learning through hints. Additionally, during high-stakes periods like mid-semester exams, ChatGPT’s advanced models have proven especially effective for tackling analytical questions. This complementary usage highlights that while Disha excels in promoting critical thinking and gradual skill development, ChatGPT is relied upon for its accuracy, speed and efficiency in resolving complex issues. 
Consequently, students benefit the most from a balanced learning approach, using human mentoring for foundational learning and complex scenarios, Disha's guided tips to build their problem-solving skills, and ChatGPT's direct answers to ensure correctness and quick problem-solving in time-constrained situations. 

\subsection{\textbf{Comparison with Other Tools}}

To evaluate the unique contributions of the Sakshm AI platform, we conducted a comparative analysis with other popular AI-driven educational platforms, including ChatGPT, LeetCode, CodeAid, and Codehelp \cite{liffiton2023codehelp}. Table \ref{tab:comparison} highlights the features that differentiate these platforms in areas critical to coding education.

\subsubsection{\textbf{AI Assistant}}
Sakshm AI stands out as the only platform with a fully integrated AI assistant that provides comprehensive support to users. Unlike other platforms such as ChatGPT and CodeAid, where AI assistance is limited as they don't have the context of the code, or LeetCode, which lacks AI features altogether, Sakshm AI ensures continuous support throughout the learning process. This integration significantly enhances its usability for learners at different skill levels.

\subsubsection{\textbf{Socratic Guidance}}
Sakshm AI excels in implementing Socratic guidance, encouraging critical thinking by providing hints and structured questions rather than direct answers. While Codehelp \cite{liffiton2023codehelp} also employs this approach, its scope is narrower compared to Sakshm AI. This is because it doesn't have the context of the user's question, test case, code, etc. All these have to be explicitly provided.  Other platforms, such as ChatGPT and LeetCode, do not utilize Socratic methods, instead offering direct answers or static content, which may not effectively foster deeper learning.

\subsubsection{\textbf{Contextual Hints}}
Contextual hints, a key feature of Sakshm AI, provide users with tailored guidance that aligns with their progress and problem-solving history. This ensures a seamless learning experience and reduces the need for repetitive queries. While CodeAid offers some contextual hints, its effectiveness is limited as the code has to be given. Also, it doesn't have the user's problem-solving history. ChatGPT, LeetCode, and Codehelp \cite{liffiton2023codehelp} lack this feature entirely, requiring users to structure their prompts manually for relevant outputs.

\subsubsection{\textbf{Prompt Friendliness}}
Sakshm AI is highly prompt-friendly, as it maintains the context of the user's code and queries, allowing for efficient and meaningful interactions. Other platforms, such as ChatGPT and LeetCode, do not have the user context, making it harder for users to construct effective prompts. CodeAid and Codehelp \cite{liffiton2023codehelp}show some level of prompt friendliness but do not match the comprehensive implementation seen in Sakshm AI, as the code has to be pasted.

\subsubsection{\textbf{Feedback on Code Quality}}
Sakshm AI provides detailed code quality feedback, focusing on efficiency, readability, correctness, and structure. It also suggests improvements and tells the topics to revise for each problem. This feature helps users develop professional coding habits. Other platforms, like ChatGPT, LeetCode, and CodeAid, offer limited feedback, primarily focused on correctness, without providing actionable insights for improvement. This positions Sakshm AI as a leader in guiding users toward writing industry-standard code.

\subsubsection{\textbf{Company-Specific Questions}}
Sakshm AI includes curated questions from top companies like Google, Flipkart, and Amazon, making it an excellent resource for interview preparation. While Codehelp \cite{liffiton2023codehelp} offers some support in this area, other platforms, such as ChatGPT and LeetCode, lack this feature or provide very limited coverage. Sakshm AI’s emphasis on company-specific questions ensures that users are better prepared for industry challenges.

\subsubsection{\textbf{Topic-Wise Practice}}
Topic-wise practice is another strong feature of Sakshm AI, allowing users to focus on specific areas of improvement. This feature is also available on LeetCode but is missing or poorly implemented (depending on the user prompt) on other platforms like ChatGPT, CodeAid, and Codehelp \cite{liffiton2023codehelp}. Sakshm AI delivers a well-rounded learning experience by combining topic-wise practice with other features.

\subsubsection{\textbf{Free Complexity Analysis}}
Sakshm AI offers free complexity analysis, enabling users to optimize their code without additional costs. While LeetCode provides similar functionality, it is restricted to premium users. Platforms like ChatGPT, CodeAid, and Codehelp lack this feature, limiting their ability to effectively teach code optimization.

\subsubsection{\textbf{Code Execution}}
Sakshm AI integrates code execution directly into its platform, eliminating the need for users to switch between tools. LeetCode also offers this functionality, but it is absent in other platforms such as ChatGPT, CodeAid, and Codehelp. This integration makes Sakshm AI a convenient and holistic platform for practicing coding.

This analysis demonstrates how Sakshm AI combines the best aspects of existing tools while introducing unique features tailored for comprehensive coding education. Its Socratic guidance, contextual hints, and detailed feedback set it apart.

\subsection{Recommendations for Sakshm AI and Future Improvements}

Based on the findings and user feedback, several recommendations can help Sakshm AI AI further enhance its platform and address existing limitations. These suggestions focus on improving user experience, expanding features, and ensuring scalability for diverse learning needs.

\subsubsection{Enhancing User Interface and Experience}
\begin{itemize}
    \item \textbf{Autosave and Stability:} Implement an autosave feature to prevent users from losing progress due to refreshes or technical glitches.
    \item \textbf{Dark Mode Improvement:} Fully implement dark mode to cover all interface sections, improving usability during extended sessions.
    \item \textbf{Customization Options:} Allow users to customize the chatbot's persona, including response length and tone, to align with individual preferences.
\end{itemize}

\subsubsection{Expanding Feature Set}
\begin{itemize}
    \item \textbf{Advanced Career-Oriented Tools:} Add features like a functional resume builder, interview preparation modules, and mock technical assessments to cater to job seekers.
    \item \textbf{Broader Language Support:} Extend programming language coverage to include popular languages like R, C\#, and SQL, enabling more learners to benefit from the platform.
    \item \textbf{Collaboration Tools:} Introduce collaborative coding spaces where users can work with peers or mentors, promoting community engagement.
\end{itemize}

\subsubsection{Adaptive Learning Enhancements}
\begin{itemize}
    \item \textbf{Dynamic Hinting:} Develop adaptive hints that adjust to a user’s skill level and time constraints, ensuring appropriate guidance without being overly simplistic or detailed.
    \item \textbf{Personalized Learning Paths:} Introduce tailored practice recommendations based on individual performance and areas of improvement.
\end{itemize}

\subsubsection{Expanding Problem Types}
\begin{itemize}
    \item \textbf{Real-World Scenarios:} Incorporate real-world problem-solving exercises and multi-concept challenges to prepare users for practical coding applications.
    \item \textbf{Advanced Topics:} Include problems in advanced areas like machine learning, cloud computing, and parallel programming to cater to experienced users.
\end{itemize}

\subsubsection{Feedback System Upgrades}
\begin{itemize}
    \item \textbf{Actionable Feedback:} Enhance feedback reports with step-by-step explanations of errors and suggestions for improvement, especially for logic and syntax issues.
    \item \textbf{Comparative Performance Analytics:} Provide insights on how a user’s performance compares with peers, including average completion time and success rates.
\end{itemize}

\subsection{Future Directions}

To advance the capabilities of LLM-based teaching agents and maximize their impact in educational contexts, the following directions should be explored:

\begin{itemize}

    \item \textbf{Interdisciplinary Applications:} Expand the scope of teaching agents to include a broader range of disciplines, such as data science, finance, and engineering, leveraging their flexible guidance capabilities for diverse domains.

    \item \textbf{Integration of Peer Learning:} Facilitate collaborative learning environments by incorporating forums or discussion boards where users can exchange insights, discuss solutions, and collectively tackle challenges, fostering a sense of community and shared growth.

    \item \textbf{Longitudinal Impact Studies:} Conduct long-term evaluations to assess the effectiveness of LLM-based teaching agents in improving user proficiency, critical thinking, and skill application, focusing on their real-world implications for professional success.

    \item \textbf{Gamification and Rewards:} Incorporate gamified elements such as badges, leaderboards, and rewards to sustain user motivation and engagement over time, encouraging consistent participation and progress.

    \item \textbf{Accessibility and Inclusivity:} Enhance the accessibility of these platforms by including features such as multilingual support, assistive technologies for differently-abled users, and localized content to cater to diverse user needs and global audiences.

    \item \textbf{Enhancing the Socratic Approach for Broader Applicability:} While the Socratic guidance method has proven effective in promoting critical thinking, it may not equally suit all users or problem types. For instance, feedback indicates that this approach is more effective for users with foundational knowledge and for moderately difficult questions. At the same time, beginners or those tackling highly complex problems may find it overwhelming.

    To address this limitation, LLM-based teaching agents could incorporate a scaffolded teaching model alongside the Socratic approach:
    \begin{itemize}
        \item \textbf{Breaking down complex problems into simpler sub-problems:} Teaching agents could guide users through smaller, manageable steps that simplify the learning process for challenging issues.
        \item \textbf{Encouraging synthesis and application:} After guiding users through sub-problems, teaching agents could prompt them to synthesize their learning and apply it to solve the original, more complex problem.
    \end{itemize}
    
\end{itemize}

By adopting these strategies, LLM-based teaching agents can provide a more inclusive and versatile learning experience, fostering critical thinking while addressing the diverse needs of learners at varying skill levels.

\section{Conclusion}\label{sec:conclusion}
This study comprehensively evaluates Sakshm AI, an AI-powered coding assistant designed to enhance problem-solving proficiency through Socratic tutoring. Unlike traditional AI-driven tutors that provide direct solutions, Sakshm AI facilitates conceptual understanding by offering guided hints and interactive feedback. To assess its efficacy, we conducted a large-scale analysis of 1,170 users, exploring engagement patterns across multiple dimensions, including problem difficulty levels, chat interactions, hourly activity distributions, and quartile-based user segmentation. Our findings reveal a strong correlation between chat-assisted learning and problem-solving efficiency and structured engagement trends, with peak usage occurring during post-school hours.

The results underscore key design principles for AI-powered educational platforms, emphasizing the need for a balanced approach that enhances independent problem-solving while offering appropriate guidance. Further advancements in question diversity, AI-driven reasoning for complex problem-solving, and user experience enhancements could significantly amplify Sakshm AI’s effectiveness. Using insights from quantitative and qualitative analyses, this research contributes to the broader discussion on AI-assisted education, providing a framework for developing scalable, intelligent tutoring systems that optimize student engagement and learning outcomes.

\bibliographystyle{ACM-Reference-Format}
\bibliography{references}

\section{Interview Questions}\label{sec:interviewProtocols}
\subsection*{Study Tools and Learning Preferences}
\begin{enumerate}
    \item Do you use any platforms, tools, or apps for studying? If yes, which ones?
    \item When you have doubts about a subject or lecture, how do you typically resolve them?
    \item Do you use AI-based tools for learning? If so, which ones and how do they help you?
\end{enumerate}

\subsection*{Sakshm Platform Experience}
\begin{enumerate}[resume]
    \item What features of the Sakshm platform do you find helpful?
    \item What aspects of the Sakshm platform could be improved?
\end{enumerate}

\subsection*{Disha (AI Assistant) Usage}
\begin{enumerate}[resume]
    \item Have you used Disha for guidance? How helpful do you find its responses?
    \item What improvements could be made to Disha to better support your learning?
    \item Have you faced any challenges using the Sakshm platform or Disha?
    \item Do you feel confident in implementing concepts after using Disha, without needing further assistance?
\end{enumerate}

\subsection*{AI and Learning Approaches}
\begin{enumerate}[resume]
    \item Do you find that AI tools like Disha enhance your ability to learn coding, or do they slow down your progress?
    \item When solving coding problems, do you prefer step-by-step guidance or direct answers? In which situations would each approach be more helpful?
    \item How does your experience with Sakshm’s AI assistant compare to other coding platforms? Which do you prefer, and why?
\end{enumerate}

\subsection*{Impact of AI on Problem Solving}
\begin{enumerate}[resume]
    \item Has using AI tools changed how you approach problem-solving compared to when you don’t use AI tools?
    \item Have you ever needed to use external resources (e.g., Google) while interacting with Disha? If so, why?
\end{enumerate}

\subsection*{Feedback and Improvements}
\begin{enumerate}[resume]
    \item Is the feedback report provided by Sakshm helpful? Is there any information you think should be added?
    \item How can the overall Sakshm platform be improved to better support your learning experience?
\end{enumerate}

\subsection*{LLM Tutor vs. Human Assistance}
\begin{enumerate}[resume]
    \item How would you compare an AI-based tutor like Disha to in-person teaching assistants or instructors in terms of approachability and accuracy?
    \item Do you think LLMs help in mastering coding skills, or do they hinder your ability to think independently?
\end{enumerate}

\section{Codebook}\label{sec:codebook}
\begin{table*}[h]
    \centering
    \begin{tabular}{lp{8cm}r}
        \toprule
        \textbf{Themes} & \textbf{Codes} & \textbf{Count} \\
        \midrule
        \textbf{Study and Coding Practices} & & \textbf{36} \\
        & Platforms for Practice & 10 \\
        & Learning Approaches and Tools & 9 \\
        & Challenges and Motivations & 11 \\
        & Visual Learning Preferences & 6 \\
        \midrule
        \textbf{Resolving Doubts and Seeking Help} & & \textbf{64} \\
        & AI Assistance & 20 \\
        & Human Support & 14 \\
        & Sakshm AI and Disha & 12 \\
        & Search and Online Resources & 18 \\
        \midrule
        \textbf{Effectiveness of Sakshm Platform} & & \textbf{59} \\
        & AI Features and Disha & 17 \\
        & Content and Coverage & 19 \\
        & Practice and Progress Tracking & 13 \\
        & User Interface and Navigation & 10 \\
        \midrule
        \textbf{Challenges in Using the Platform} & & \textbf{74} \\
        & AI Guidance Issues & 13 \\
        & Feedback on Assessment and Practice & 12 \\
        & Limited Variety and Real-World Relevance & 19 \\
        & Technical and Performance Challenges & 16 \\
        & User Experience and Interface Improvements & 14 \\
        \midrule
        \textbf{Feedback and Reporting in the Platform} & & \textbf{45} \\
        & Detail and Depth of Feedback & 15 \\
        & Grading and Scoring Mechanism & 16 \\
        & Feedback for Advanced Learning & 14 \\
        \midrule
        \textbf{Impact of Socratic Guidance by AI Chatbot} & & \textbf{47} \\
        & Applicability in Learning Scenarios & 11 \\
        & Encouragement of Independent Thinking & 15 \\
        & Insights from Socratic Feedback & 12 \\
        & Limitations of the Socratic Approach & 9 \\
        \midrule
        \textbf{Balancing AI and Human Assistance} & & \textbf{45} \\
        & Accessibility vs. Personalization & 12 \\
        & Combining AI and Human Support & 13 \\
        & Efficiency vs. Depth of Learning & 11 \\
        & Independent Thinking and Reliance & 9 \\
        \midrule
        \textbf{Comparison with Other Tools} & & \textbf{57} \\
        & Promotes Critical Thinking and Non-Direct Answers & 15 \\
        & Key Features and Integration of Disha & 17 \\
        & Interface and Usability & 14 \\
        & Comparison with Other Platforms & 11 \\
        \bottomrule
    \end{tabular}
    \caption{Summary of Themes and Codes with Count}
    \label{tab:themes_codes}
\end{table*}


\end{document}